%% file: main.tex
\documentclass[journal]{IEEEtran}%
\usepackage{xcolor}

\usepackage{amsthm}
\theoremstyle{plain}

  \theoremstyle{plain}
  
\providecommand{\corollaryname}{Corollary}
\providecommand{\theoremname}{Theorem}

\usepackage[style=ieee,
            isbn=false,
            doi=false,
            url=false]
            {biblatex}
\ifCLASSINFOpdf
  \usepackage[pdftex]{graphicx}
  \graphicspath{{img/}}
  \DeclareGraphicsExtensions{.pdf,.jpeg,.png}
\else
   \usepackage[dvips]{graphicx}
  \graphicspath{{img/}}
   \DeclareGraphicsExtensions{.eps}
\fi
\usepackage[cmex10]{amsmath}
\usepackage{amssymb}
\ifCLASSOPTIONcompsoc
 \usepackage[caption=false,font=normalsize,labelfont=sf,textfont=sf]{subfig}
\else
 \usepackage[caption=false,font=footnotesize]{subfig}
\fi
\usepackage{url}
\usepackage{lipsum}
\usepackage{soul}
\usepackage{xcolor}
\usepackage{colortbl}
\usepackage{enumitem}
\usepackage{algorithm}
\usepackage{algpseudocode}
\usepackage[colorlinks=true,bookmarks=false,citecolor=blue,urlcolor=blue]{hyperref}

\renewcommand{\ul}{}

\definecolor{Gray}{gray}{0.85}
\newcommand{\cfg}[1][]{\emph{configuration} #1}
\input{acronyms}
\input{commands-optcomm}
\input{commands-nft}
\input{commands-ml}
\begin{document}
%\title{End-to-end optimization of a coherent communication system using solitonic pulse shaping}
%\title{End-to-end optimization of a coherent communication system: an auto-encoder trained over the split-step Fourier method}
%\title{End-to-end optimization of coherent optical communications trained over the split-step Fourier method}
%\title{Autoencoder for coherent optical communications trained over the split-step Fourier method} % Backup
% \title{End-to-end optimization of a coherent optical communication system trained over the split-step Fourier method} 
\title{End-to-end optimization of coherent optical communications over the split-step Fourier method guided by the nonlinear Fourier transform theory} 

%\author{XXXX
%\thanks{YYYY}% <-this % stops a space

\author{Simone Gaiarin, Francesco Da Ros, Rasmus T. Jones, and Darko Zibar
  \thanks{Manuscript received April 1st, 2020; }%
  \thanks{S. Gaiarin, F. Da Ros, and D. Zibar are with the Department of Photonics Engineering, Technical University of Denmark, Kongens Lyngby, 2800 Denmark, e-mail: \{simga,fdro,dazi\}@fotonik.dtu.dk}% <-this % stops a space
  \thanks{R. T. Jones was with  with the Department of Photonics Engineering, Technical University of Denmark, Kongens Lyngby, 2800 Denmark. He is now with Oticon, Sm{\o}rum, 2765 Denmark, email: rajo@fotonik.dtu.dk}%
 %\thanks{Manuscript received February XX, 2017; revised XX , 2017.}% <-this % stops a space
}
% and the end of the author line. i.e. if you had this:

\markboth{Submitted to IEEE Journal of Lightwave Technology}{}% {}% <-this % stops a space
%{Da Ros \MakeLowercase{\textit{et al.}}: On-Chip Optical Phase Conjugation...}% <-this % stops a space



\maketitle

\begin{abstract}
Optimizing modulation and detection strategies for a given channel is critical to maximize the throughput of a communication system. Such an optimization can be easily carried out analytically for channels that admit closed-form analytical models. However, this task becomes extremely challenging for nonlinear dispersive channels such as the optical fiber. End-to-end optimization through \acp{AE} can be applied to define symbol-to-waveform (modulation) and waveform-to-symbol (detection) mappings, but so far it has been mainly shown for systems relying on approximate channel models. Here, for the first time, we propose an \ac{AE} scheme applied to the full optical channel described by the \ac{NLSE}.  Transmitter  and receiver  are  jointly  optimized  through  the  \ac{SSFM}  which  accurately  models  an  optical  fiber. 
% In order to simplify and guide the optimization on the modulation side, in this first numerical analysis,  whereas the detection is performed by a \ac{NN}, the symbol-to-waveform mapping is aided by the \ac{NFT} theory.
In this first numerical analysis,  the detection is performed by a \ac{NN}, whereas the symbol-to-waveform mapping is aided by the \ac{NFT} theory in order to simplify and guide the optimization on the modulation side.
This proof-of-concept \ac{AE} scheme is thus benchmarked against a \ul{manually-optimized} \ac{NFT}-based system and a three-fold increase in achievable distance (from 2000 to 6640 km) is demonstrated.
\end{abstract}
\acresetall


% \fi
% \IEEEpeerreviewmaketitle

\section{Introduction} % THIS IS A COPY OF THE INTRO
\label{sec:Intro}

\IEEEPARstart{O}{ptical} communication systems are striving to maximize the achievable information rate distance product. In order to achieve this goal, optimizing the signal constellation (e.g. constellation shaping~\cite{MetodiJLT16,BuchaliJLT16,RennerJLT17,JonesECOC2018}) may not be sufficient and modulation and detection strategies need to be jointly improved. Whereas for simple transmission channels, such as \ac{AWGN} channels, it is possible to analytically derive optimal modulation and detection strategies, such an optimization is particularly challenging for the nonlinear optical channel. As a closed-form expression to describe the signal propagation through an optical fiber is not available, current research relies on approximate analytical or numerical models achieving only a limited degree of accuracy \cite{oliari2020regular}. Even following this direction though, no definitive answer to optimal modulation and detection strategies is know.
In order to address this challenges, full end-to-end learning through \acp{AE}, which does not require closed-form channel models,  has been proposed~\cite{OShea2018}. The first proposal analyzing an \ac{AWGN} channel~\cite{OShea2018} was followed by a number of works focusing on the optical fiber channel, e.g. \cite{Li2018,JonesArxiv2018,JonesECOC2018,Karanov2018, Karanov2019a}, however, all relying on approximate channel models. 
A general \ac{AE} model of a coherent communication system (complex signal transmission) targets at using the full nonlinear dispersive channel model and replacing the transmitter and the receiver with \acp{NN}. After the training phase, this model can provide two key functions: the optimal encoding of symbols onto time-domain waveforms resilient to the optical channel impairments; the decoding of the received waveforms to recover the transmitted symbols. An ideal \ac{AE} should therefore both be trained on an accurate channel model, as well as provide symbol-to-waveform and waveform-to-symbol transformation. 
%Current results in literature have shown accurate detection (waveform-to-symbol mapping) schemes but have yet to address challenge of modulation (symbol-to-waveform mapping) in the context of an accurate channel model.

In this work, we extend our initial proposal of \cite{GaiarinCLEO20}, discussing the first numerical analysis of an \ac{AE} scheme for coherent communication making use of the \ac{SSFM} to accurately model the optical fiber channel. However, in order to restrict the space of all possible modulation strategies, in this work, we demonstrate our proposed method by guiding the optimization through the mathematical theory of the \ac{NFT}~\cite{hasegawa1995solitons,Ablowitz2004a}. Our proposed \ac{AE} scheme jointly optimizes an \ac{NFT}-aided transmitter with an \ac{NN}-based receiver over the channel modeled by solving the \ac{NLSE} through the \ac{SSFM}. % An \iac{NFT} transmitter is chosen as the \ac{NFT} theory describes how to construct waveforms that are spectrally invariant to the impact of dispersion and Kerr nonlinearity for a channel modeled by the lossless \ac{NLSE}~\cite{TuritsynOptica17,LeNatPhot2017,Gaiarin2018,DongPTL15,DaRosJLT19}.
This choice results in pulse-shapes (solitons) which do not suffer from significant pulse broadening over propagation, therefore enabling a small (low-complexity) and memoryless (1-symbol in input) \ac{NN} to be used at the receiver. Due to the system similarity with \ac{NFDM} systems, the proposed scheme is benchmarked against standard \ac{NFDM} transmission (\ac{NFT} transmitter and receiver). This proof-of-concept demonstration shows more than three times extension in transmission reach \ul{compared to a manually-optimized }\ac{NFDM} \ul{ system}. The 2000-km achieved by standard \ac{NFDM} transmission are extended to more than 6000~km through the end-to-end optimization.

The remaining of the paper is organized as follows. In Section~\ref{sec:e2e-comm-system} the concept of \ac{AE} is reviewed, discussing its application to communication systems and positioning this work within the state-of-the-art on this topic. In Section~\ref{sec:num-setup} the numerical setup is presented, including a brief review of the fundamentals of \ac{NFT}. The specific system model and its implementation are also described in detail. In Section~\ref{sec:training} the training procedure used is detailed and key practical trade-offs are identified. The key transmitter and receiver parameters that have been optimized are introduced, and the different optimization cases are presented. In Section~\ref{sec:results} the main results of this work are reported, starting with the results of the optimization and the training performance and following with the performance of the system during the testing phase. The different optimized transmission schemes are compared against each other and benchmarked against a conventional \ac{NFDM} system. Finally, the conclusions are summarized in Section~\ref{sec:conclusions}.

\section{End-to-end communication systems}
\label{sec:e2e-comm-system}

\begin{figure}[t]
  \centering
  \includegraphics[width=.7\linewidth]{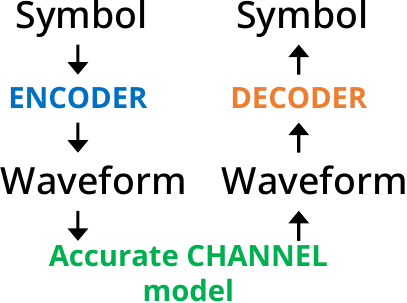}
  \caption{Ideal \acl{AE} model for an optical communication system.}
  \label{fig:AE}
\end{figure}

\begin{figure*}[ht]
  \centering
  \includegraphics[width=.95\linewidth]{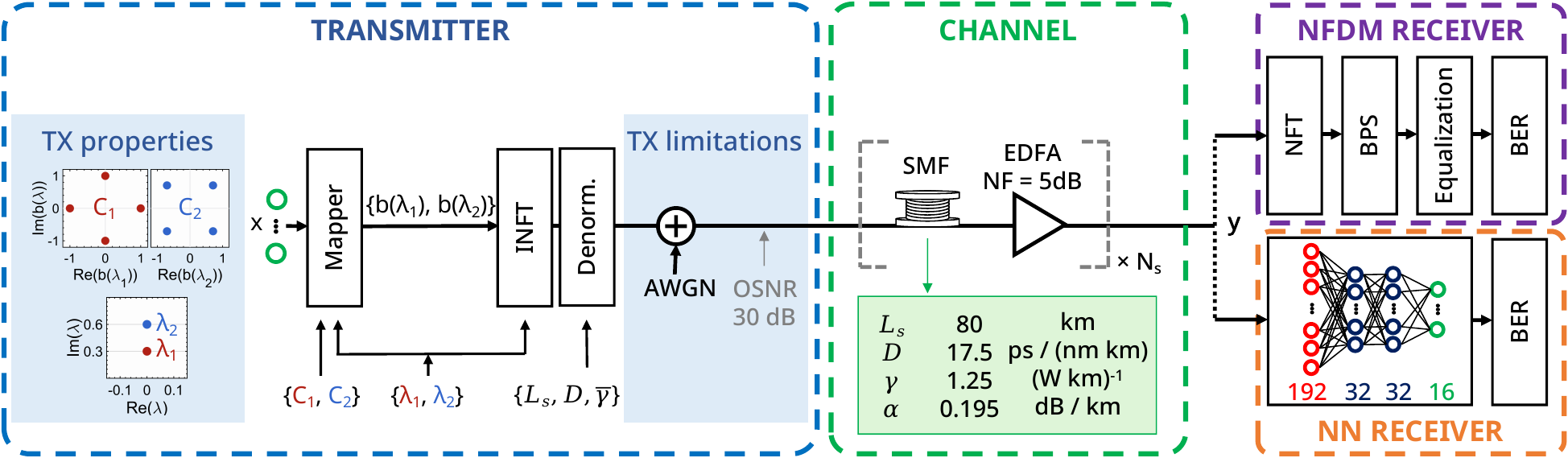}
  \caption{Setup of the optical communication system used for both the end-to-end optimization of the system parameters and the \ac{BER} performance evaluation. }
  \label{fig:setup}
\end{figure*}

A communication system is composed of three main blocks: a transmitter, a channel, and a
receiver. The goal of the system is to faithfully reproduce the information entered at
its input to its output, i.e. to estimate the transmitted symbol sequence with the lowest error probability. 
The structure of the communication system resembles that of \iac{AE} \cite{OShea2018}, as shown in \figurename~\ref{fig:AE}. In its general definition, \iac{AE} consists of two key transformations: an encoder function that maps the input data to a \emph{code}, i.e. an encoded version of the data,  and a decoder function that tries to reconstruct the original data from the code \cite{goodfellow2016deep}. According to the specific problem and context, different \ac{AE} architectures can be chosen to generate codes optimized according to different metrics, e.g. lower dimensionality compared to the input data, or robustness to noise and distortion sources.
For the case of a communication system, the transmitter, i.e. the encoder, maps the information (symbols) to a time-domain  waveform, i.e. the code, that is transmitted over the channel. The channel is the source of noise and distortion. The receiver, i.e. the decoder, is responsible to recover the original data symbols from the corrupted version of the code, which is the waveform after the transmission. The goal for training the transmitter and the receiver of this communication system is to minimize the error probability of the transmitted symbols, and it is achieved by generating transmitter waveforms that are resilient to the noise and distortions introduced by the channel.
% Usually, the encoder and decoder of \iac{AE} are implemented as \acp{NN}, given that these can theoretically approximate any function thus offering the most unrestricted space of solutions for the \emph{code} \cite{goodfellow2016deep}. However, such a large solution space for the  may make the training of the \ac{AE} particularly challenging in some situations.

The ideal \ac{AE} for optical coherent communication systems needs to consider the full nonlinear dispersive channel model and can be constructed by replacing the transmitter and the receiver with \acp{NN}. Indeed a sufficiently large \ac{NN} can theoretically approximate any function, including the one that generates this optimal set of waveforms~\cite{musumeci2018overview}. Nonetheless, to achieve this final goal the correct optimization strategy and \ac{NN} architecture must be devised, and this is challenging to do in practice, especially for non-trivial channels such as the optical fiber.
 Therefore, the preliminary demonstrations reported for optical communications have considered simplified version of the overall structure of \figurename~\ref{fig:AE}. All the works reported consider approximate channel models such as 
 %\ac{AWGN} channels~\cite{OShea2018},  Rayleigh channels~\cite{Stark2019}, 
 a simplified memoryless nonlinear channel model~\cite{Li2018}, a dispersive linear fiber channel model \cite{Karanov2018, Karanov2019a},  and perturbative models of the nonlinear fiber channels~\cite{JonesArxiv2018,JonesECOC2018}. %Others performed the full optimization of the time-domain waveforms entering a dispersive linear fiber channel \cite{Karanov2018, Karanov2019a}, however only focusing on \ac{IM} and disregarding fiber nonlinearity. 
% Additionally, only the work from~ \cite{Karanov2018, Karanov2019a} discussed full encoding, i.e. from symbols to waveforms, whereas the other reports focused on mapping from bit/symbol to complex constellation, and assumed conventional pulse shaping.
 End-to-end learning for the full nonlinear dispersive fiber channel, i.e. not relying on perturbative/approximate models, has yet to be reported.
In order to approach the desired system shown in \figurename~\ref{fig:AE}, in this work,  the \ac{AE} considers the \ac{SSFM} to numerically implement the physical channel. The encoder and decoder are therefore trained over a highly accurate representation of the channel, moving one step closer to the final goal of Fig.~\ref{fig:AE}.

Concerning the encoder, however, of the available literature, only the work from~\cite{Karanov2018, Karanov2019a} discussed full encoding, i.e. from symbols to waveforms, whereas the other reports focused on mapping from bit/symbol to complex constellations, and assumed conventional pulse shaping. A fully blind \ac{AE} consisting of two \acp{NN} incurs in the challenge of a vast and complex optimization landscape. 
%At the beginning of the training, a randomly initialized \ac{NN} transmitter would generate a random waveform that after the propagation through the channel may ill-condition the backpropagation algorithm thus making the training likely to fail. 
% It needs to be noted that, whereas impressive, the results of~ \cite{Karanov2018, Karanov2019a} consider intensity-modulation therefore halving the dimensionality of the problem.
In ~\cite{Karanov2018,Karanov2019a}, the optimization was aided by considering intensity-only modulation and, therefore, halving the dimensionality of the problem.
Here, in order to avoid this challenge worsened by the even more complex optimization landscape introduced by the accurate channel model, the receiver/decoder is replaced by \iac{NN} whereas the transmitter/encoder is implemented as a conventional \ac{NFDM} transmitter with trainable parameters. The \ac{NFT} theory helps in guiding the optimization of the encoder by liming the solutions space of the time-domain waveforms generated by the transmitter.  In particular, the \ac{NFDM} transmitter is constrained to solitonic pulse-shapes, which are exact analytical solutions of the \ac{NLSE} in absence of loss, and have been shown to provide appropriate pulse shapes also for practical transmissions with losses \cite{TuritsynOptica17, Gaiarin2018}. Moreover, given that solitons do not disperse with the transmission distance (the pulse broadening can be kept minimal also in the presence of losses \cite{hasegawa1995solitons}), they are not subject to strong \ac{ISI} and can be detected with a low-complexity memoryless receiver as the \ac{NN} considered in this work. 

\section{Numerical setup}
\label{sec:num-setup}
The complete simulation setup used for both the end-to-end optimization of the communication system and for the
performance evaluation is depicted in \figurename~\ref{fig:setup} and its
parameters are summarized in \tablename~\ref{tab:sim_params}. The same figure also shows the standard \ac{NFDM} receiver that has been used to compute a benchmark test performance.
The setup was implemented in Tensorflow to perform the optimization of the parameters (training phase, Section~\ref{sec:training}) while the performances of the system using the optimal parameters have been estimated using our existing MATLAB implementation of the setup (test phase, Section~\ref{sec:results}). 
% If the reviewers ask us why we did the test in MATLAB, is because we had to do it in any case for the NFT, so we decided to use the same setup to compare the two fairly. If they want the test in TF we can compute the test accuracy in SER.
In the following sub-sections, after a brief introduction to the \ac{NFT} theory, transmitter, channel, and receivers (both \ac{NN}-based and conventional for \ac{NFDM}) of \figurename~\ref{fig:setup} are described in detail.

% \hl{In Section~}\ref{subsec:trainable_parameters}\hl{ it will be discussed how to relax the soliton amplitude-duration condition to reduce the limits on the solution space.}

\subsection{\Acl{NFT}}
\label{sec:nft}

The complex field envelope $\flds = \fldl$ of a
single polarization signal propagating in \iac{SSMF} with losses evolves according to the
\ac{NLSE}
%%%% NLSE %%%%
\begin{equation}\label{eq:NLSE}
    \pdv{\flds}{\ssp}= -\dfrac{\loss}{2}\flds -\iunit\dfrac{\dispersion}{2}\pdv[2]{\flds}{\ttm}+\iunit\gamma|\flds|^2\flds.
\end{equation}
where $\ttm$ is the retarded time, $\ssp$ the distance, \ul{$\loss$ the attenuation coefficient}, $\dispersion$ the \ac{GVD}, and $\nonlinfact$ the Kerr nonlinear coefficient of the fiber.
The direct and inverse \ac{NFT} transforms are defined for a lossless and
noiseless \ac{NLSE} ~\cite{Ablowitz2004a}. To abstract from the specific channel
parameters this \ac{NLSE} is usually presented in its normalized form that for the
anomalous dispersion regime \mbox{($\dispersion<0$)} becomes
%%%% MANAKOV SYSTEM NORMALIZED %%%%
\begin{equation}\label{eq:NNLSE}
    \iunit\pdv{\nflds}{\nssp}= \pdv[2]{\nflds}{t}+2|\nflds|^2\nflds\\
\end{equation}
where $\nttm$ is the normalized retarded time, and $\nssp$ the normalized distance. This
equation is derived from \eqref{eq:NLSE} ignoring the loss term and through the change of variables
%\eqref{eq:change_of_variables_NLSE}, here reported for convenience
%%%% CHANGE OF VARIABLES %%%%
\begin{equation}\label{eq:normalization}
   \nfld = \dfrac{\fld}{\sqrt{P}}, \hspace{0.7cm}
   \nttm = \dfrac{\ttm}{T_0}, \hspace{0.7cm}
   \nssp = -\dfrac{\ssp}{\nftL},
\end{equation}
with $P = |\dispersion|/(\nonlinfact T_0^2)$,
$\nftL = 2 T_0^2/|\dispersion|$, and $T_0$ a free normalization parameter.

The direct \ac{NFT} maps a time-domain waveform to a so-called nonlinear spectrum.
When the time-domain waveforms are solitons, the nonlinear spectrum is said to be
discrete and it is composed of a set of complex eigenvalues (nonlinear frequencies) $\eig[i]$ with associated complex scattering coefficients $\nfta[i]$, $\nftb[i]$ with $\nfta[i] = 0$~\cite{Gaiarin2018}.

The nonlinear spectrum can be modulated and used to carry information.
A discrete \ac{NFDM} communication system works as follows: the data is encoded onto the discrete nonlinear spectrum that is then mapped to a solitonic time-waveform through an \ac{INFT} operation. The waveform is transmitted through the nonlinear fiber channel. At the receiver, the waveform is mapped back to a nonlinear spectrum (direct \ac{NFT}) to retrieve the encoded data. Specific details of the \ac{NFDM} communication system are given in the following sections.

\begin{table}[b]
  \centering \caption{Simulation parameters}
  \setlength\tabcolsep{3 pt}
  \begin{tabular}{llcc}
    \hline
    &Parameter name & Parameter & Value                                \\
    \hline
    Transmitter&\# Polarizations   &  & 1 \\
    &Symbol rate    & & 1 \gbaud \\
    &Oversampling    & & 96 \\
    &\ac{NFT} free normalization factor& $\To$ & 47 ps \\
    &\Acl{LPA} $\nonlinfactlpa$& $\nonlinfactlpa[LPA]$ & 0.34 \\
    Channel&Span length &$\lspan$ & 80 km \\
    &Dispersion parameter &$D$ &  17.5 ps / (nm km)\\
    &Nonlinear coefficient &$\nonlinfact$ &  1.25 (W km)$^{-1}$\\
    &Attenuation &$\loss$ & 0.195 dB/km \\
    &EDFA noise figure&  & 5 dB \\
    Receiver \ac{NN}&\# nodes in input layer  && \ul{192}\\
    &Activation functions && SeLu\\
    &\# hidden layers &&  2\\
    &\# nodes per hidden layer  && 32\\
    &\# nodes in output layer  && 16\\
    \hline
  \end{tabular}
  \label{tab:sim_params}
\end{table}

\subsection{Transmitter}
\label{sec:tx}

The transmitter is a standard single-polarization \ac{NFDM} transmitter with the same structure as the one reported in \cite{JonesPTL2018} and experimentally validated in~\cite{GaiarinJLT2020}.

In details, a sequence of uniform random
%one-hot-encoded
symbols \mbox{$\txsymbonehot \in \symbalphabet =
%\{\mathbf{e}_i~|~i = 1, \dots, M = 16 \}$, where $\mathbf{e}_i$ is equal to 1 at row $i$ and 0 elsewhere,
\{ i~|~i = 1, \dots, M \}$}, with $M = 16$ the size of the chosen symbol alphabet,
is generated.  Each symbol is mapped onto a
discrete nonlinear spectrum consisting of two discrete eigenvalues $\eig[i]$ and their associated
complex scattering coefficients $\nftb[i]$, $i=1,2$, which are independently modulated using
4-\ac{PSK} as illustrated in \figurename~\ref{fig:setup}. The set of values \{$\eig[i], \nftb[i]$\}, $i=1,2$ constitutes \iac{NFDM} symbol that carries a total of four information bits, similarly to two parallel 4-\ac{PSK} channels, at the symbol rate of 1 \gbaud{}. Note that the cardinality of the alphabet, as well as the number of eigenvalues, have not been optimized. The values have been chosen mainly to simplify the comparison with existing literature on \ac{NFDM} systems~\cite{Gaiarin2018,JonesPTL2018,DaRosJLT19,LeNatPhot2017,DongPTL15,GaiarinJLT2020}.
\ul{The choice of b-modulation stems from its better noise resilience, as discussed in}~\cite{wahls2017generation}.
%, i.e. \mbox{2 bit/symbol/eigenvalue}.

An \ac{INFT} implemented with \iac{DT} \cite{matveev1991darboux} maps each \ac{NFDM} symbol into a time-domain waveform $\nfld$ with 96 samples-per-symbol and solitonic pulse shape.

In order to obtain an optical field, $\fld$, matched to the \ac{NLSE} in \eqref{eq:NLSE}, the waveform $\nfld$ needs to be de-normalized (Denorm. block in \figurename~\ref{fig:setup}) using
\eqref{eq:normalization} with the parameters reported in \tablename~\ref{tab:sim_params}. This last operation ensures that the optical field, which carries the information, is matched to a lossless optical channel. Given that the fiber loss is not accounted for by the \ac{NFT} theory, which relies on \eqref{eq:NNLSE}, the obtained waveform is not perfectly matched to the actual channel constituted of lossy \ac{SMF} spans interleaved by \acp{EDFA}.
It is possible to obtain a better match by performing the de-normalization in
\eqref{eq:normalization} replacing the fiber
nonlinear coefficient $\nonlinfact$ with a different value $\nonlinfactlpa$. In a standard \ac{NFDM} communication system this
value is usually set to the one provided by the \acf{LPA} approximation \cite{hasegawa1995solitons}, which for
the channel considered in this work is $\nonlinfactlpa[LPA] = 0.34$.
%$\nonlinfactlpa[LPA] = \nonlinfact(G_a - 1) / \log(G_a)$ and $G_a = \exp(-\loss \lspan)$ .
% It should be noted that the parameter $\nonlinfactlpa$ affects only the amplitude de-normalization parameter $P$, thus, given that the symbol period is fixed at 1~ns, changing $\nonlinfactlpa$ is equivalent to changing the launch power of the signal $\fldl$.

After de-normalization, the field $\fld$ is used to ideally modulate a laser, thus disregarding transmitter impairments such as laser phase noise and distortions introduced by the \ac{MZM}. The obtained signal is loaded with \ac{AWGN} noise to limit its \ac{OSNR} to 30~dB and finally transmitted over the channel. 
% The transmitter parameters trained by the \ac{AE} will be discussed in Section~\ref{sec:training}.

\subsection{Channel}
\label{sec:ch}

The channel is composed of $\nspans$ spans of $\lspan = 80$-km long
\ac{SSMF} interleaved by \acp{EDFA} (5-dB noise figure) used to compensate for the fiber loss. The full set of channel parameters are reported in \tablename{}~\ref{tab:sim_params}.

The signal propagation is numerically simulated using the well-know \ac{SSFM} \cite{AgrawalBook}.
Being the \ac{SSFM} a composition of basic differentiable mathematical operations, it is
possible to numerically compute the gradient propagation from the output of the channel to its
input, thus enabling an optimization of the performance of the communication system in terms of \ac{BER} across the channel. The \ac{SSFM} in this work is implemented in Tensorflow, and thanks to the automatic differentiation capability of the Tensorflow framework \cite{baydin2017automatic}, the gradient is automatically computed. One limitation of a Tensorflow implementation of the channel is that it is not possible to use the variations of the
\ac{SSFM} using an adaptive step size. Indeed Tensorflow requires that the computational graph that implements the channel must be created before the computation happens. However, using a predetermined constant step size within the \ac{SSFM} yields very slow and unpractical computations for the batch size and oversampling rate used in this work, especially for long transmission distances. As a trade-off between computing time, \ac{SSFM} precision and Tensorflow requirements, the \ac{SSFM} is implemented with \ul{80 fixed step sizes per span}, that increase logarithmically along the fiber length \cite{BoscoPTL00}. 
This choice of step sizes guarantees that at each step of the propagation the signal undergoes a maximum nonlinear phase rotation of 0.01 degrees when the simulation is performed with the
reference configuration (see Section~\ref{sec:training}).
\ul{ Additionally, the simulation bandwidth of 96-GHz }\ref{sec:training}\ul{ is sufficient to ensure that the} \ac{SSFM} \ul{provides an accurate solution to the} \ac{NLSE} \ul{ describing the optical signal propagation.}

% The signal coming from the channel is sliced in non-overlapping blocks $\rxsymbwave$ of 96 samples (each corresponding to a single \ac{NFDM} symbol) that are sequentially fed to the receiver, either \ac{NFT}- or \ac{NN}-based.

\subsection{\Acl{NFT} receiver}
\label{sec:nft_receiver}
% THIS SENTENCE IS ALREADY IN THE RESULTS, AND HERE IT IS REFERRING TO SOMETHING COMING LATER
% The \ac{AE}-optimized transmitter parameters have also been tested using an \ac{NFT} receiver. Whereas the receiver considered in the optimization was a \ac{NN} receiver, a standard \ac{NFT} receiver provides a benchmark for the BER performance.

The \ac{NFT} receiver is a simplified version of the one in \cite{Gaiarin2018PTL}. In
particular, a bandpass filter with a bandwidth of 20~GHz has been used to filter the
out-of-band noise prior to the nonlinear spectrum computation with the \ac{NFT}. The detected discrete
nonlinear spectrum was processed with a \ac{BPS} carrier phase estimation algorithm to compensate for the phase rotation introduced by the nonlinear domain transfer function of the channel over the $\nftb[i]$ \cite{hasegawa1995solitons,Gaiarin2018PTL}. The compensated spectrum was finally equalized with \iac{LMMSE} equalizer~\cite{Gaiarin2018PTL} prior to computing the \ac{BER}.

\subsection{\Acl{NN} receiver}
\label{sec:nn_receiver}

The signal coming from the channel is sliced in non-overlapping blocks $\rxsymbwave$ of 96 samples (each corresponding to a single \ac{NFDM} symbol) that are sequentially fed to the receiver, which consists of a feed-forward \ac{NN} for multi-class classification with $M = 16$ output classes corresponding to all the possible transmitted symbols. \ul{As the }\ac{NN} \ul{is implemented as a real-valued network, the input layer consists of 192~nodes, considering the real and the imaginary parts of each sample as separate inputs.}
The \ac{NN} takes as input the sequence $\rxsymbwave$ and provides at the output node $i$ the probability $p(\txsymbonehot{}_i|\rxsymbwave)$ of having transmitted the symbol $\txsymbonehot{}_i \in \symbalphabet$ given the received vector $\rxsymbwave$. The activation function of the hidden layers is \iac{SELU} \cite{Klambauer2017} while the output layer uses a softmax activation function \cite{goodfellow2016deep}. The \ac{NN} weights are initialized using the Glorot algorithm \cite{glorot2010understanding}\ul{, and optimized within the end-to-end training discussed in Section~}\ref{sec:training}.
The receiver \ac{NN} parameters are reported in \tablename~\ref{tab:sim_params}.
An argmax operation on the \ac{NN} output probabilities is used to perform a hard decision that provides the estimated transmitted symbol $\rxsymbonehot$. The detected symbols are finally used to compute the \ac{BER}.
Remark that the use of a memoryless (1-input symbols) receiver is enabled by the choice of an \ac{NFT}-aided transmitter, which in turn results in transmitted waveforms (solitons) not affected by significant pulse broadening. 
% The \ac{BER} of the system is finally estimated assuming a gray coding on the individual $\nftb[i]$ constellations.

\section{End-to-end training}
\label{sec:training}

\begin{figure}[hb]
  \centering
  \includegraphics[width=.95\linewidth]{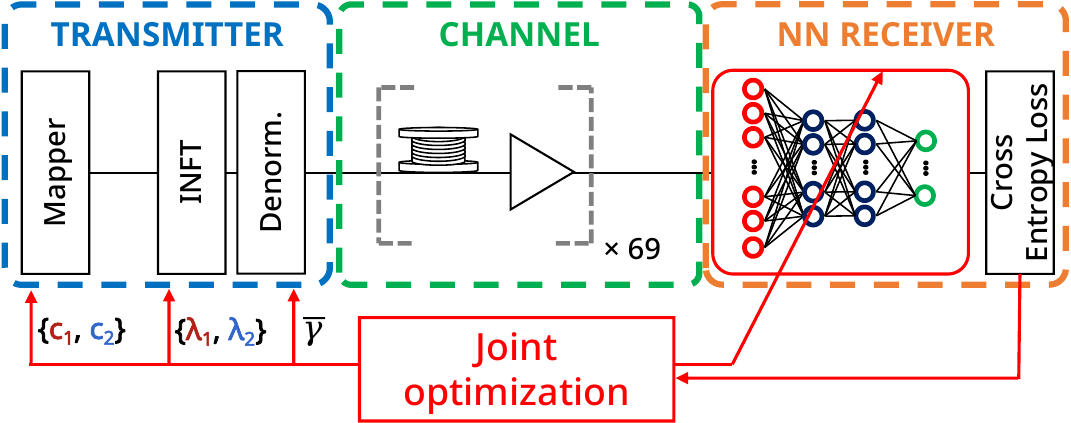}
  \caption{Setup used for the training phase highlighting the transmitter and receiver parameters jointly optimized by the \ac{AE}.}
  \label{fig:setup_training}
\end{figure}

The goal of training the \ac{AE} is to maximize the probability that the output symbol
$\rxsymbonehot$ of the communication system is equal to the input symbol
$\txsymbonehot$~\cite{OShea2018,khan2019optical}.
% (?This is equivalent to maximize the mutual information between the
% transmitted and received symbols \cite{Li2018}?).
The training consists of iteratively
varying a set of trainable parameters and evaluating the performance of the system in
terms of cross-entropy loss \cite{goodfellow2016deep}. The details on the trainable parameters and the
training process are given in the next two sections
and the whole process is highlighted in \figurename{}~\ref{fig:setup_training}.
\ul{Note that, due to the data processing inequality, the joint optimization of transmitter and receiver through an} \ac{AE} \ul{approach, will theoretically yield equal or better performance compared to independent optimization of transmitter and receiver. The disadvantage of an independent optimization of the different blocks of a communication system is already introduced in }~\cite{Karanov2018}.

\subsection{Trainable parameters}
\label{subsec:trainable_parameters}
The parameters of the transmitter chosen for the optimization are the imaginary part of
the purely-imaginary eigenvalues $\eig[i]$, and the radius $r_i$ and phase $\cphase[i]$ of the two \ac{PSK}
\mbox{$\nftb[i]$-constellations}:
\begin{equation}
C_i = r_i \exp \left( \iunit \left( k
\frac{\pi}{2} + \cphase[i] \right ) \right),~ k=0,1,2,3;~i=1,2.
\label{eq:C_i}
\end{equation}
Unlike a classical coherent modulation where the impulse response of a fixed pulse shaping filter is linearly modulated by the amplitude and phase of the transmitted symbols, the \ac{INFT} nonlinearly maps the symbols-eigenvalue pairs to a time-domain waveform that is theoretically optimal for transmission over the nonlinear channel modeled by the lossless \ac{NLSE}.
Therefore, by constraining the transmitter pulse shape through the \ac{INFT}, it is the nonlinear relation between symbols and waveforms which is kept fixed. This is in contrast with previous literature~\cite{JonesArxiv2018,JonesECOC2018} where the pulse shape is kept fixed regardless of the symbols optimization. This choice, whereas not reaching the target \ac{AE} scheme of \figurename~\ref{fig:AE}, it moves one step close in that direction.
The components of the nonlinear spectrum affect the generated waveform $\nfld$ pulse shape in the following way \cite{Ablowitz2004a}:
\begin{enumerate}[label=\alph*)]
\item The imaginary part of the eigenvalues controls the energy $E$ of the waveform according to
\begin{equation}\label{eq:parseval}
E =  4 \sum_{i=1}^2 \Im(\eig[i]),
\end{equation}
and, at the same time, the waveform duration. Indeed the amplitude and
duration of soliton pulses are inversely related. The energy of the waveform is only dependent on the $\eig[i]$s and it is
independent on the $\nftb[i]$-constellations.
\item The radii $r_i$ of the constellations control the temporal position of the two
components constituting the soliton waveform and thus their temporal overlap/separation.
\item The difference $\cphasediff$ between the constellation phases $\cphase[i]$ governs
the shape, and thus the bandwidth, of the waveform. The absolute phase of each $C_i$ is not particularly relevant. A constant phase offset of both $C_i$s maps to the same phase offset in the time-domain waveform \cite{Yousefi2014}.
\end{enumerate}

\begin{table}[h]
  \centering \caption{Transmitter trainable parameters ($\nonlinfactlpa$ optimized  in  $\nonlinfactlpa[E2E]$-scenario, $\nonlinfactlpa$ kept fixed to $\nonlinfactlpa[LPA]$ otherwise.)}
  \begin{tabular}{ccccc}
    \hline
    Configuration & $\eig[1], \eig[2]$& $\cphase[1], \cphase[2]$      & $\cradius[1], \cradius[2]$               & $\trainparams$ \\
          &                           &                           &                           & \\
    \hline
    0     & $0.3\iunit,  0.6\iunit$   & $0, 0.25 \pi$             & 1, 1                      & $[\nonlinfactlpa, \mathbf{w}_{NN}]$\\
    1     & $0.3\iunit,  0.6\iunit$   & \cellcolor{Gray}trained & \cellcolor{Gray}trained& $[C_i, \nonlinfactlpa, \mathbf{w}_{NN}]$\\
    2     & \cellcolor{Gray}trained & $0, 0.25 \pi$             & 1, 1                      & $[\lambda_i,, \nonlinfactlpa \mathbf{w}_{NN}]$\\
    3     & \cellcolor{Gray}trained & \cellcolor{Gray}trained & \cellcolor{Gray}trained& $[\lambda_i, C_i, \nonlinfactlpa, \mathbf{w}_{NN}]$ \\
    \hline
  \end{tabular}
  \label{tab:configurations}
\end{table}

To study the individual effects of these transmitter parameters on the performance of
the communication system, four different training \emph{configurations} are considered. In each configuration, a subset of
these parameters is trained, while the remaining transmitter parameters are kept constant. For each of the configurations considered, the weights and biases (one bias per layer) of the \ac{NN} that constitutes the receiver, denoted with $\nntrainparams$, are trained. For each configuration, the set of trained parameters $\trainparams$ and the static values used for the parameters that are not trained are reported in
\tablename~\ref{tab:configurations}. The static values are the same as those used in \cite{JonesPTL2018}.
In the \cfg{0}, none of the transmitter parameters are optimized.
This configuration
is used as a benchmark to compare the performance of the other configurations.
In \cfg{1}, only the constellations, i.e. radii and phases, are trained, in \cfg{2} only the eigenvalues, and finally, in \cfg{3} all parameters, constellations, and eigenvalues are fully trainable.

The choice of guiding the encoder optimization through the \ac{NFT} allows to restrict the solution space that
the \ac{AE} has to search, but it results in enforcing a strict constraint on the
waveform amplitude-duration relation \cite{Ablowitz2004a}. For example, the \ac{AE} can
decrease the imaginary part of the eigenvalues, thus decreasing the amplitude and average power of the
waveform while extending its temporal duration. The temporal broadening is however restricted by the \ac{NFDM} symbol temporal slot of 1 ns. By further decreasing the amplitude, and thus further broadening the waveforms, it will spread beyond the available symbol slot yielding a negative impact on the system performance.
This condition can be relaxed by making the variable $\nonlinfactlpa$ a trainable parameter.

% FIXME
It should be noted that the parameter $\nonlinfactlpa$ affects only the amplitude de-normalization parameter $P$, thus, given that the symbol period is fixed at 1~ns, changing $\nonlinfactlpa$ is equivalent to changing the launch power of the signal $\fldl$.

By training $\nonlinfactlpa$, any launch power can be set for a given transmitted waveform shape. Whereas this slightly deviates from the strict NFT theory, it gives the \ac{AE} one additional degree of freedom to improve the overall system performance. It is therefore interesting to compare the two scenarios to understand how strictly the \ac{NFT} theory can be applied to a lossy channel while aiming to improve transmission performance. The four different configurations of
\tablename~\ref{tab:configurations} have been optimized for both scenarios:
$\nonlinfactlpa$
is set equal to the value provided by the
\ac{LPA} approximation explained in Section~\ref{sec:tx} ($\nonlinfactlpa =
\nonlinfactlpa[LPA]$) and $\nonlinfactlpa$
 is optimized jointly with the other trainable variables ($\nonlinfactlpa =
\nonlinfactlpa[E2E]$).

\subsection{Training}

\input{algorithms/ae_training}

\begin{figure*}[t]
    \centering
    \subfloat[Power factor unoptimized ($\nonlinfactlpa =
\hat{\gamma}_{LPA}$) ]{\label{fig:const_lpa}
        \includegraphics[height=.39\linewidth]{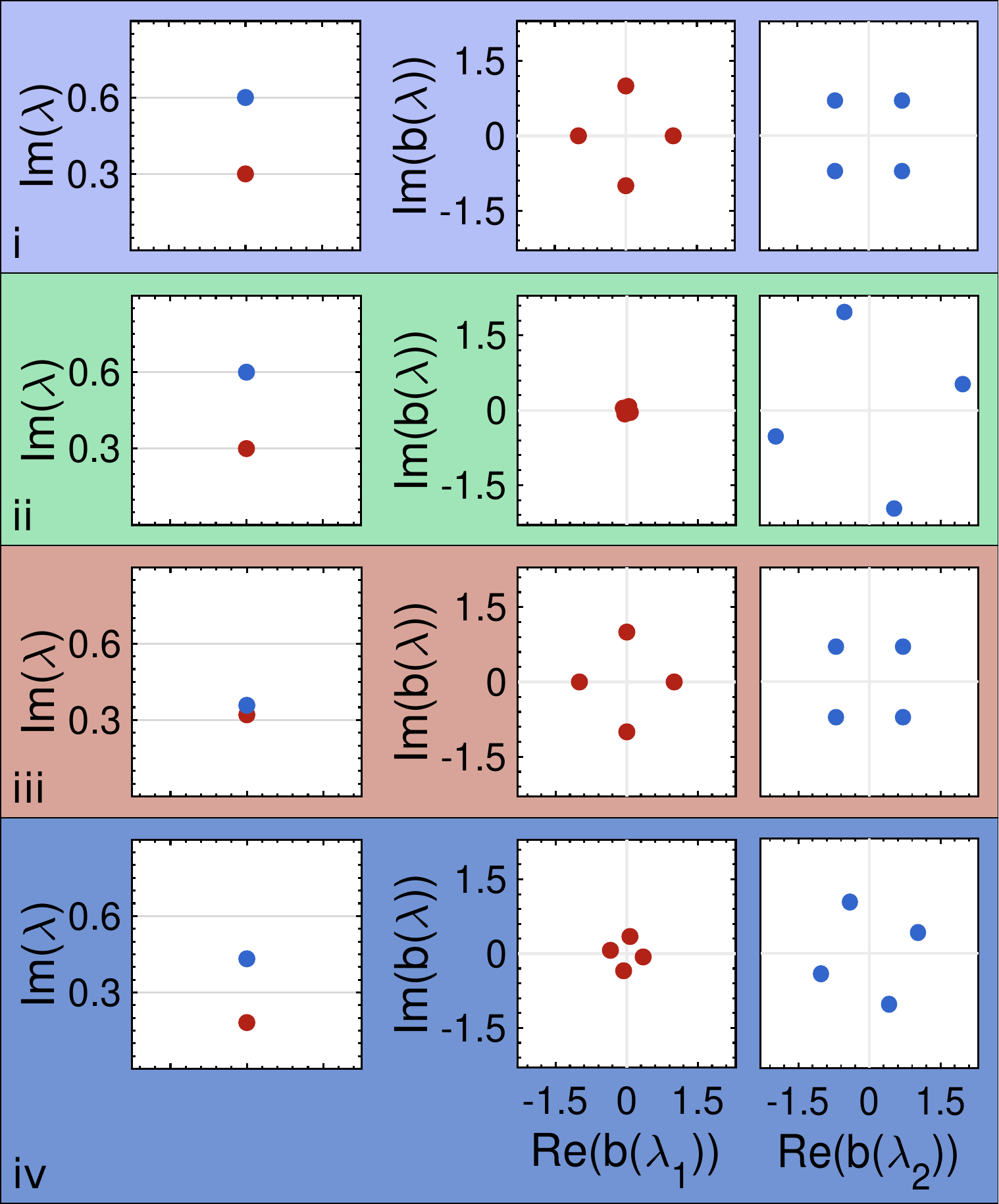}
    }
    \subfloat[Power factor optimized ($\nonlinfactlpa =
\hat{\gamma}_{E2E}$)]{\label{fig:const_e2e}
        \includegraphics[height=.39\linewidth]{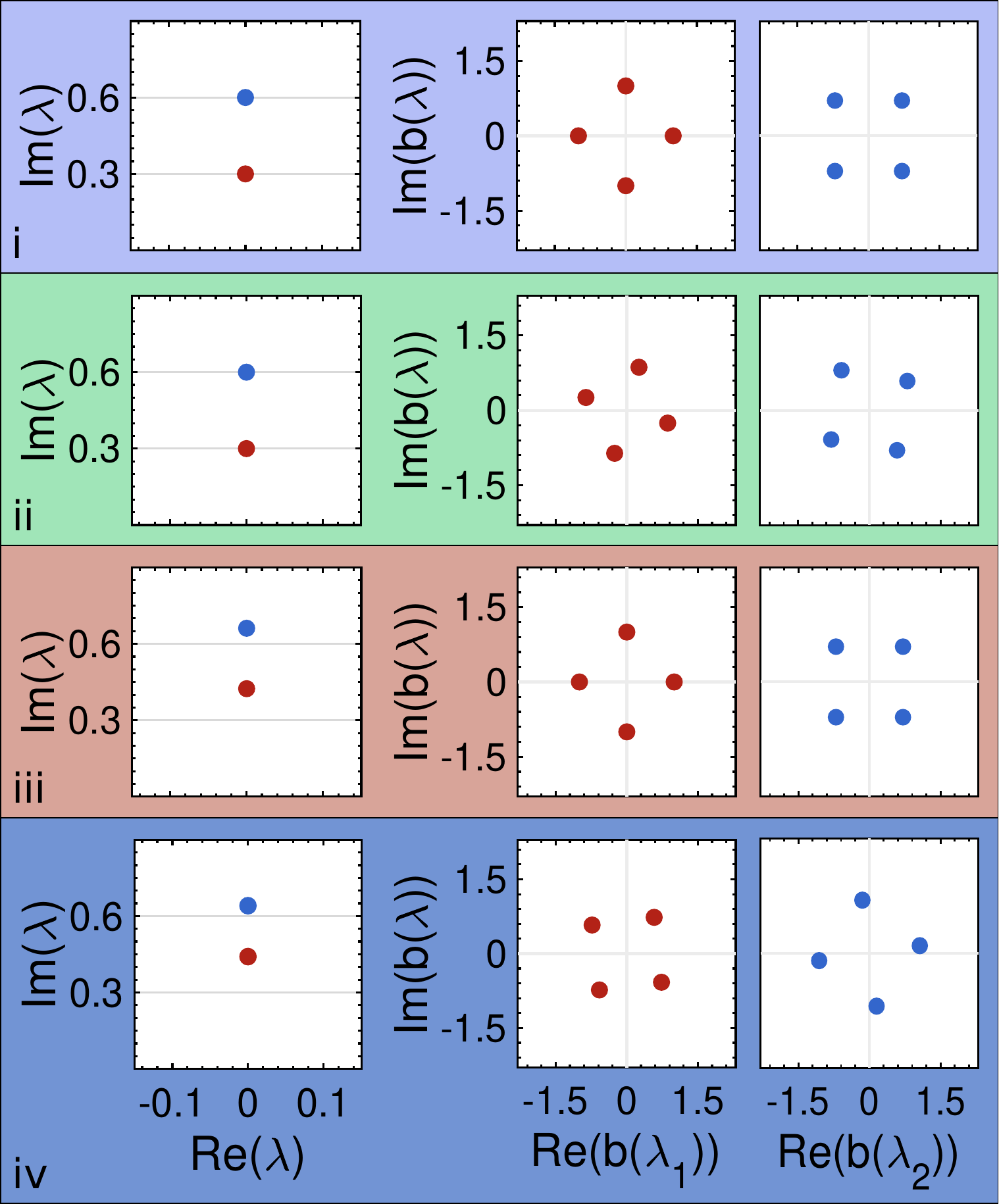}
    }
    \subfloat[Transmitted waveforms]{\label{fig:tx_waveforms}
        \includegraphics[height=.39\linewidth]{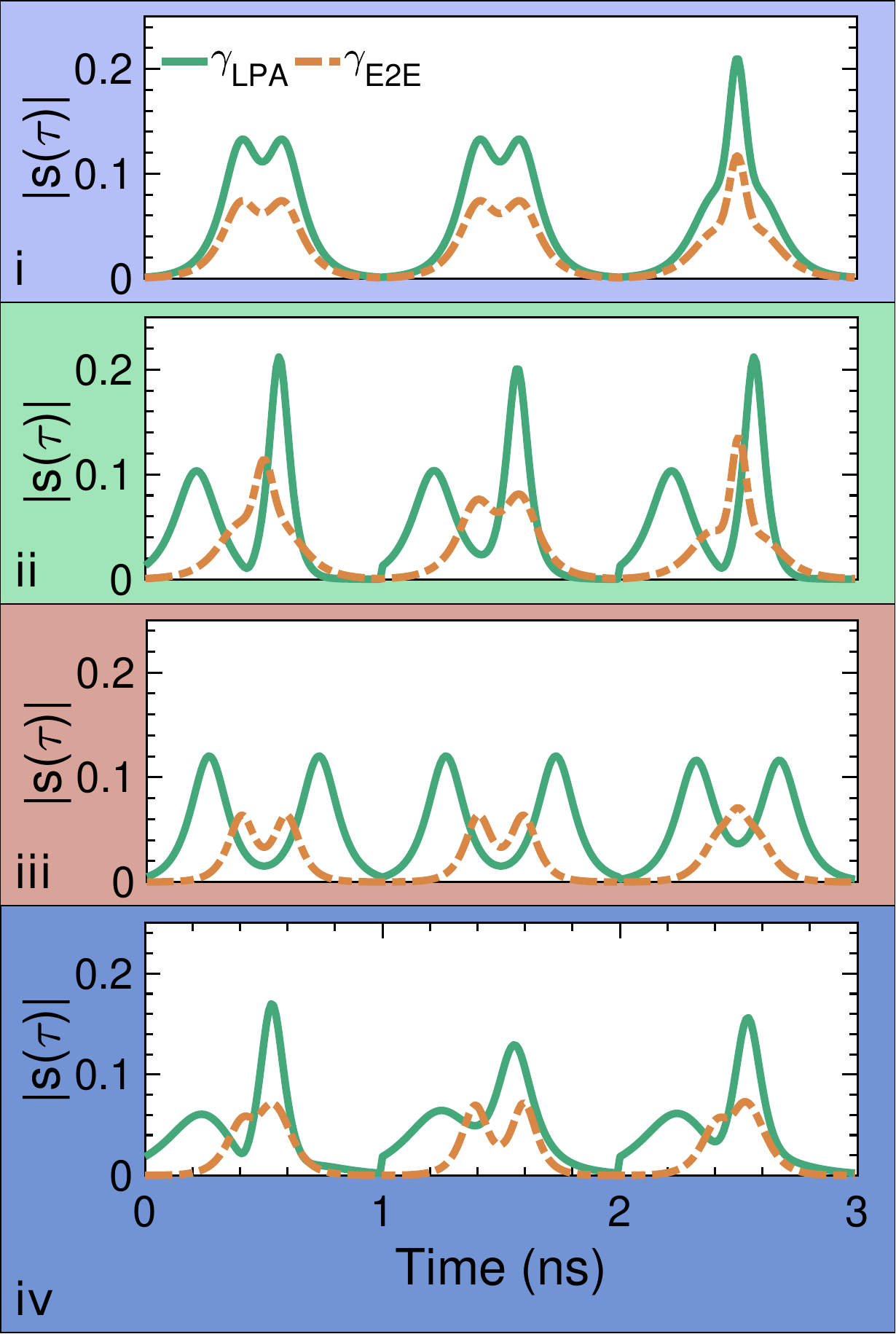}
    }
    \caption{\emph{Trained transmitter parameters} -  Eigenvalues and constellations for the four configurations (i) to (iv) and scenario $\nonlinfactlpa =
\hat{\gamma}_{LPA}$ (a) and $\nonlinfactlpa =
\hat{\gamma}_{E2E}$ (b). Transmitted waveforms for $\nonlinfactlpa =
\hat{\gamma}_{LPA}$ (solid) and $\nonlinfactlpa =
\hat{\gamma}_{E2E}$ (dashed) for the four configurations (i) to (iv). }
    \label{fig:const_and_waveforms}
\end{figure*}

The \ac{AE} optimization procedure is reported in  Algorithm~\ref{alg:ae_training}.
The transmission distance was fixed at $69 \times 80$~km.
This distance has been chosen so that for all the configurations the accuracy of the \ac{AE} during the training is not too close to saturation. This, in turn, results in \ac{BER} values for the testing phase which are neither too low (and thus challenging to count accurately), nor too high, overall allowing for a non-trivial comparison of the performance of the different configurations.
% , and permits a more accurate optimization procedure.

The \ac{AE} is trained using the Adam optimizer with Nesterov gradient \cite{dozat2016incorporating}. The
optimization is run for a total of \mbox{$\ntrainiter = 6400$}
% 16(nBatch) * 350(evaluations)
iterations, a fixed value that was observed to ensure the
convergence of all the $2~\times$ four training configurations considered.

In each
iteration, a batch of \ul{B =} 64 symbols
% --we could probably fit more, at least twice this number--
$\txsymbonehot \in \symbalphabet$ is generated. This batch size is the
maximum size for which the entire \ac{AE} network (including the \ac{SSFM}) fits into the
% --16 GB--
available memory of \ul{our}
\ac{GPU} using the memory saving gradient technique \cite{Chen2016, MemorySavingGradient}. During the forward propagation of one
training iteration, the results of each operation throughout the whole communication model - including the results of each step of the \ac{SSFM} - need to be saved, to be used later during the backpropagation that computes the gradient of the loss function.
This implies large memory requirements that limit the training batch size, which, in turn, causes a noisy estimation of the gradient at each training iteration.
\ul{Being able to increase the batch size by increasing the available computing memory, may lead to improved performance or shorter training time}~\cite{OShea2018}.
\ul{A systematic complexity analysis of the proposed approach is beyond the scope of this work. Nevertheless, the current implementation is particularly time-consuming within the training phase whereas compares favorably against our implementation of the} \ac{NFT} \ul{ receiver in the testing phase.}

The learning rate was tuned according to a step decay schedule that sets it to $\lr = 0.01$ in the first 1600 iterations, $\lr = 0.003$ for the following 2400 iterations, and $\lr = 0.001$ for the remaining 2400 iterations. This scheduling allowed a quick convergence in the initial part of the training and a fine-tuning of the performance on the final part.

The architecture and hyper-parameters of the receiver \ac{NN} have been kept fixed to provide a fair comparison among the different transmitter configurations. The chosen \ac{NN} architecture yielded a sufficiently small network size while still guaranteeing good detection performances. A fine optimization of the receiver hyper-parameters for the optimal transmitter configuration would potentially improve the communication system performance, but doing it is beyond the scope of this work.

As the \acp{EDFA} noise in the channel is randomly generated at each training iteration, no identical batch is seen more than once by the \ac{AE} during the training, even though the transmitted symbols may be the same due to the finite batch size (online learning). This prevents overfitting problems that may instead arise when using a training dataset of fixed and limited size that is reused multiple times during the training process (batch learning). \ul{This choice allows avoiding a separate cross-validation loss analysis to monitor eventual overfitting.} 

\section{Simulation results}
\label{sec:results}
In this section, the results of the \ac{AE} training and testing are discussed. In Sections~\ref{subsec:optParam} and \ref{subsec:AEperf}, the trained transmitter parameters and the training performance are reported, whereas Sections~\ref{subsec:results_nft} and \ref{subsec:results_nn} show the performance under testing for an \ac{NFT}- and \ac{NN}-based receiver, respectively.

\subsection{\ac{AE}-trained transmitter parameters}
\label{subsec:optParam}
% \begin{figure*}[t]
%   \centering
%   \includegraphics[width=.95\linewidth]{constellations_and_waveforms}
%   \caption{Constellations and waveforms}
%   \label{fig:constellations}
% \end{figure*}

The values of the transmitter trainable parameters found by the end-to-end optimization of the communication system are reported in
\tablename~\ref{tab:const_lpa_false} for $\nonlinfactlpa = \nonlinfactlpa[LPA]$, and
in \tablename~\ref{tab:const_lpa_true} for $\nonlinfactlpa = \nonlinfactlpa[E2E]$.
The tables also report the average power of the transmitted waveforms generated using those parameters.
These optimized transmitter parameters (eigenvalues and $\nftb[i]$-constellations)
and the resulting time-domain waveforms are shown in \figurename~\ref{fig:const_and_waveforms}  for \subref{fig:const_lpa} $\nonlinfactlpa = \nonlinfactlpa[LPA]$ and \subref{fig:const_e2e}
$\nonlinfactlpa = \nonlinfactlpa[E2E]$.

%Starting from the case where
%the transmitted power is only determined by the imaginary part of the eigenvalues according to \eqref{eq:parseval}
Starting from the ($\nonlinfactlpa = \nonlinfactlpa[LPA]$) scenario, a qualitative analysis of
the optimized constellations (\figurename~\ref{fig:const_lpa}) indicates
that the \ac{AE} optimizes the eigenvalues (configurations 2 and 3) by decreasing their imaginary part. As the transmitted power in this scenario is only determined by the imaginary part of the eigenvalues, decreasing the eigenvalues is equivalent to reducing the average power of the waveforms and extending their temporal duration.
Observing the corresponding waveforms (dashed curves in \figurename~\ref{fig:tx_waveforms}-(iii, iv)), the
imaginary parts of the eigenvalues are decreased up to the point where the waveforms would not fit
any more within the symbol slot of 1~ns.
% The fact that the optimization reduces the signal power indicates that the system % performance at the optimization distance are mainly limited by nonlinear effects and not % by the limited \ac{OSNR}, so that a power reduction improves the performance.
For the case where only the $\nftb[i]$-constellations can be optimized (configuration
1), the constellations found have a relative phase difference of $0.25\pi$. More importantly, the radius of the first constellation ($\cradius[1]$) is decreased during the optimization whereas the radius of the second constellation is increased ($\cradius[2]$).
These radius values cause the time waveform to have the first (second) solitonic
component respectively delayed (anticipated) with respect to the center of the time window,
as can be seen in \figurename~\ref{fig:tx_waveforms}-(iii). Interestingly, the \ac{AE} optimization (both in terms of phase difference and radii) converges to a set of constellations very similar to that reported in \cite{Gaiarin2018} where the
constellations were manually optimized to \ul{maximize performance by minimizing} the \ac{PAPR} of the generated waveforms,\ul{ and further investigated in} \cite{GaiarinJLT2020}. \ul{Note that the transmission system in }\cite{Gaiarin2018,GaiarinJLT2020}\ul{ was equivalent, i.e. 80-km spans with lumped }\ac{EDFA}\ul{ amplification. In the following, this system will therefore be considered as the reference NFDM system for benchmarking}
Finally, in \cfg{3} the radii of the constellations are tuned by the \ac{AE}
similarly to the previous case, but given that now the imaginary part of the eigenvalues
is reduced, enlarging the temporal duration of the soliton, the value of the radii of the
constellations is closer to 1 compared to the \mbox{\cfg{1}}. A deviation from unitary radii would quite rapidly result in waveforms extending beyond the allowed symbol slot.

When the constraint of the amplitude-duration of the soliton is removed
($\nonlinfactlpa = \nonlinfactlpa[E2E]$), the strategy of the \ac{AE} to reduce the average power of the generated waveforms is even more evident. As shown by comparing \tablename~\ref{tab:const_lpa_true} to \tablename~\ref{tab:const_lpa_false}, $\nonlinfactlpa$ is optimized in order to significantly reduce the power launched into the channel.
As the power can be reduced through $\nonlinfactlpa$, for this scenario, the eigenvalues are not varied as heavily as for the previous scenario, and their imaginary part is rather slightly increased instead.
Additionally, the radii of the constellations are less affected and they keep values close to unity even for configurations 1 and 3, leading to a lower separation in time between the two solitonic components. Finally, the relative phase rotations, instead, converge to similar values as for the scenario of $\nonlinfactlpa = \nonlinfactlpa[LPA]$ as can be inferred by comparing the values
reported in \tablename~\ref{tab:const_lpa_true} with those in
\tablename~\ref{tab:const_lpa_false}.

\begin{table}[t]
  \centering 
  \caption[caption]{Optimized parameters ($\nonlinfactlpa = \nonlinfactlpa[LPA]$).\\The highlighted values are the results of the training.}
  \setlength\tabcolsep{3.5 pt}
  \begin{tabular}{cccccc}
    \hline
    Configuration & $\eig[1], \eig[2]$                       & $\cphasediff$               & $\cradius[1], \cradius[2]$               & $\nonlinfactlpa[LPA]$  & Power  \\
          &                                          &                             &                            && (dBm)      \\
    \hline
    0     & $0.3\iunit,  0.6\iunit$                  & $0.25 \pi$                  & 1, 1                       &0.34& 7.03       \\
    1     & $0.3\iunit,  0.6\iunit$                  & \cellcolor{Gray}$0.25 \pi$ & \cellcolor{Gray}0.08, 2.03 &0.34& 7.03       \\
    2     & \cellcolor{Gray}$0.33\iunit, 0.37\iunit$ & $0.25 \pi$                  & 1, 1                       &0.34& 5.97        \\
    3     & \cellcolor{Gray}$0.18\iunit, 0.43\iunit$ & \cellcolor{Gray}$0.32 \pi$  & \cellcolor{Gray}0.35, 1.10 &0.34& 5.34        \\
    \hline
  \end{tabular}
  \label{tab:const_lpa_false}
\end{table}
% BW: 11.32, 12.72, 7.87, 9.65

\begin{table}[t]
  \centering \caption[caption]{Optimized parameters ($\nonlinfactlpa = \nonlinfactlpa[E2E]$)\\The highlighted values are the results of the training.}
  \setlength\tabcolsep{3.5 pt}
  \begin{tabular}{cccccc}
    \hline
    Configuration & $\eig[1], \eig[2]$                     & $\Delta \cphase$               & $\cradius[1], \cradius[2]$                 & $\nonlinfactlpa[E2E]$& Power \\
          &                                          &                             &                            && (dBm)      \\
    \hline
    0     & $0.3\iunit,  0.6\iunit$                  & $0.25 \pi$                  & 1, 1                       &\cellcolor{Gray}1.09& 1.96       \\
    1     & $0.3\iunit,  0.6\iunit$                  & \cellcolor{Gray}$0.29 \pi$ & \cellcolor{Gray}0.90, 0.99 &\cellcolor{Gray}0.97& 2.46       \\
    2     & \cellcolor{Gray}$0.42\iunit, 0.66\iunit$ & $0.25 \pi$                  & 1, 1                       &\cellcolor{Gray}2.41& -0.67      \\
    3     & \cellcolor{Gray}$0.44\iunit, 0.64\iunit$ & \cellcolor{Gray}$0.26 \pi$ & \cellcolor{Gray}0.93, 1.07 &\cellcolor{Gray}2.04& 0.04       \\
    \hline
  \end{tabular}
  \label{tab:const_lpa_true}
\end{table}
% BW: 11.32, 11.64, 11.38, 11.12
% delta phi values:
% -0.21 -> 0.29
% -0.24 -> 0.26

\subsection{\Acl{AE} training performance}
\label{subsec:AEperf}
\begin{figure*}[htb]
  \centering
  \subfloat[Power factor unoptimized ($\nonlinfactlpa =
\hat{\gamma}_{LPA}$) ]{\label{fig:ae_history_false}
    \includegraphics[width=.45\linewidth]{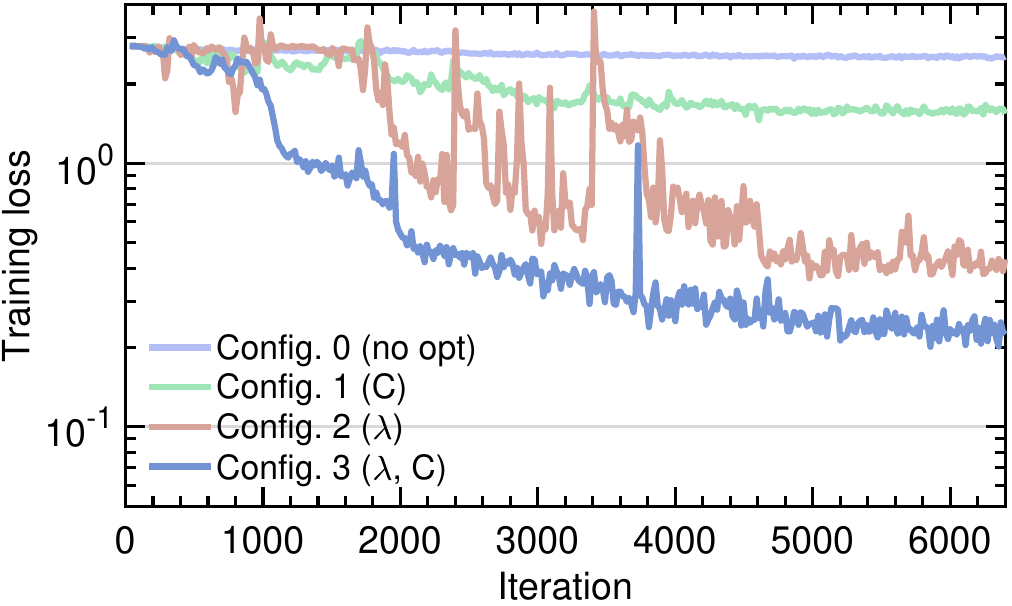}
  }
  \subfloat[Power factor optimized ($\nonlinfactlpa =
\hat{\gamma}_{E2E}$)]{\label{fig:ae_history_true}
    \includegraphics[width=.45\linewidth]{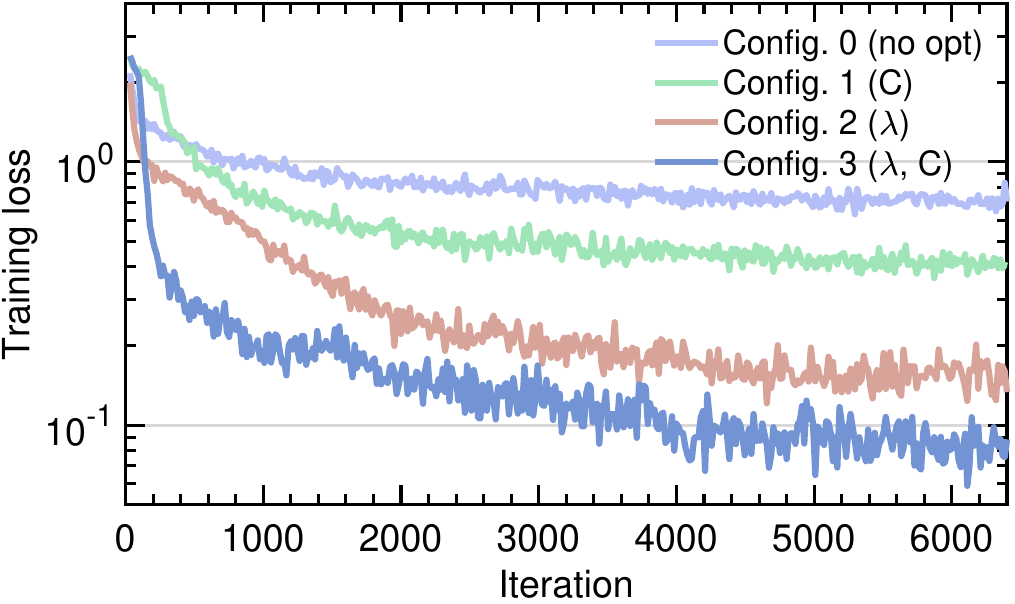}
  }
  \caption{\emph{Training performance} Cross-entropy loss (training loss) during the training process for the four configurations within the two scenarios.}
  \label{fig:ae_history}
\end{figure*}

Looking at the cross-entropy loss curves as a function of the end-to-end training iteration in \figurename~\ref{fig:ae_history}, it is possible to observe that in the scenario when $\nonlinfactlpa$ is not trained ( $\nonlinfactlpa[LPA]$, \figurename~\ref{fig:ae_history_false}), the loss curve  does not converge to values as low as those for the scenario when $\nonlinfactlpa$ is optimized (scenario $\nonlinfactlpa[E2E]$, \figurename~\ref{fig:ae_history_true}).
Moreover, in the first scenario the loss curves
do not decrease monotonically nor as quickly and smoothly as in the second scenario. In
order to reduce the instability in the convergence, the learning rate of the \ac{AE} was
optimized. Nevertheless, the curves are still unstable and the performance reported in
\figurename~\ref{fig:ae_history_false} shows the lowest losses achieved. The unstable
behavior and the poorer performance obtained suggest that the loss function for the $\nonlinfactlpa = \nonlinfactlpa[LPA]$ scenario is not smooth and
contains multiple local minima that slow down and hinder the overall optimization.

Focusing on \cfg{1} for both scenarios though, the optimization brings improvement in terms of loss compared to the reference case (\cfg{0}), but not as great as
optimizing the imaginary part of the eigenvalues (\cfg{2}). By optimizing
both degrees of freedom (\cfg{3}) it is possible to further reduce the cross-entropy loss but only minimally compared to \cfg{2}, i.e. eigenvalue-only optimization. 
This shows that the optimization of the eigenvalues is critical for the performance of the system.
These general trends during the training phase are well aligned with the results achieved during the testing phase and shown in Section~\ref{subsec:results_nn}.

\subsection{\Acl{NFT} receiver \ac{BER} performance}
\label{subsec:results_nft}

\begin{figure}[ht]
  \centering
  \includegraphics[height=.6\linewidth]{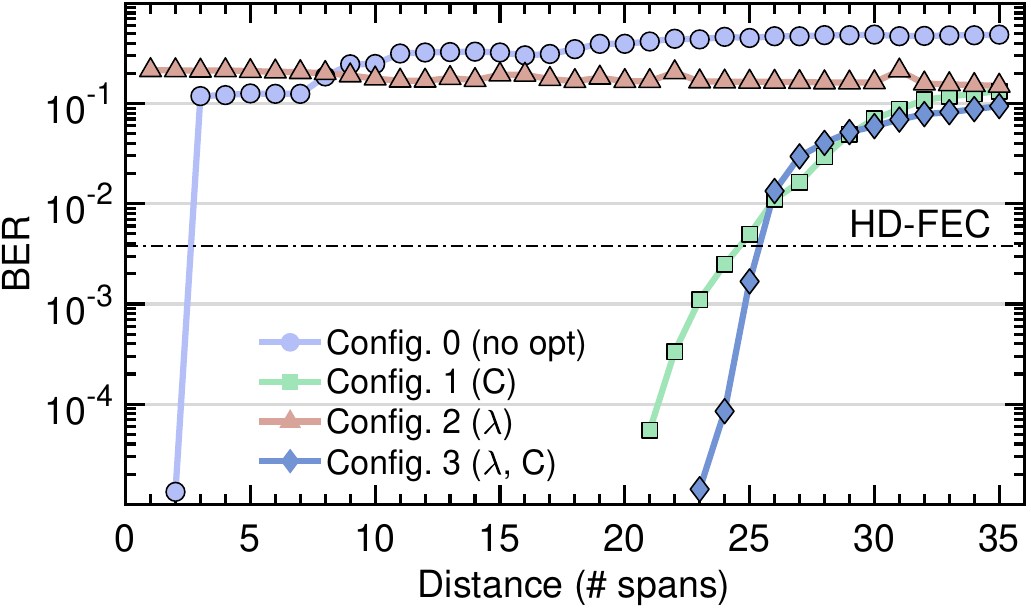}
  \caption{\emph{Test performance \ac{NFT} receiver} - \ac{BER} as a function of the transmission distance using the \ac{AE}-optimized transmitter with $\nonlinfactlpa = \nonlinfactlpa[LPA]$ and the \ac{NFT} receiver.}
  \label{fig:ber_vs_distance_nft}
\end{figure}

Although the transmitter parameters are optimized for \iac{NN} receiver,
and thus are not necessarily optimal for \iac{NFT} receiver, the performance of a conventional \ac{NFDM} system (\ac{NFT} transmitter + \ac{NFT} receiver) are presented here. The counted
\acp{BER} as a function of the  transmission distance for the four configurations are shown in \figurename~\ref{fig:ber_vs_distance_nft}.
Only the scenario where $\nonlinfactlpa = \nonlinfactlpa[LPA]$ has been
considered, as once the power of the waveform is not matched to the channel according to the NFT theory, a conventional \ac{NFDM} receiver fails to correctly demodulate the received signal. The \ac{BER} of the \ac{NFT} receiver has been computed using $5 \times 10^5$ symbols.

For \cfg{0} (unoptimized transmitter), the \ac{BER} degrades rapidly and reaches values above the
\ac{HD-FEC} threshold (\ac{BER} = $3.8 \times 10^{-3}$) already after a $3 \times 80$~km transmission. This is consistent with the results from \cite{JonesPTL2018}, and it is due to the presence of fiber loss and long fiber spans. As the power varies significantly over the fiber length, the channel strongly deviates from the lossless fiber channel over which the \ac{NFT} is defined, even considering the \ac{LPA} approximation.
When only the eigenvalues are optimized (\cfg{2}) the
receiver cannot decode the data for any of the transmission distances. This is expected
as the two eigenvalues are almost overlapped making it extremely challenging for the receiver to discriminate between them.
% BER = 0.25 vs. 0.5
When the $\nftb[i]$-constellations are optimized (\emph{configurations} 1 and 3) the \ac{BER} performance improves drastically
compared to the previous configurations, and it is possible to reach a transmission
distance  of $24 \times 80$~km spans and $25 \times 80$~km, respectively, still considering \iac{HD-FEC} \ac{BER} target.

In the future, to fully test the limits of the \ac{NFT} system, the end-to-end training can be performed using the \ac{NFT} receiver. This, however, requires to implement the \ac{NFT} transform in Tensorflow using only non-adaptive algorithms. This limitation poses a challenge because it excludes the possibility to use some commonly used algorithms for locating the eigenvalues such as the search methods \cite{Yousefi2014p2}.

\begin{figure*}[t]
    \centering
    \subfloat[Power factor unoptimized ($\nonlinfactlpa =
\hat{\gamma}_{LPA}$) ]{\label{fig:ber_vs_distance_false}
        \includegraphics[width=.49\linewidth]{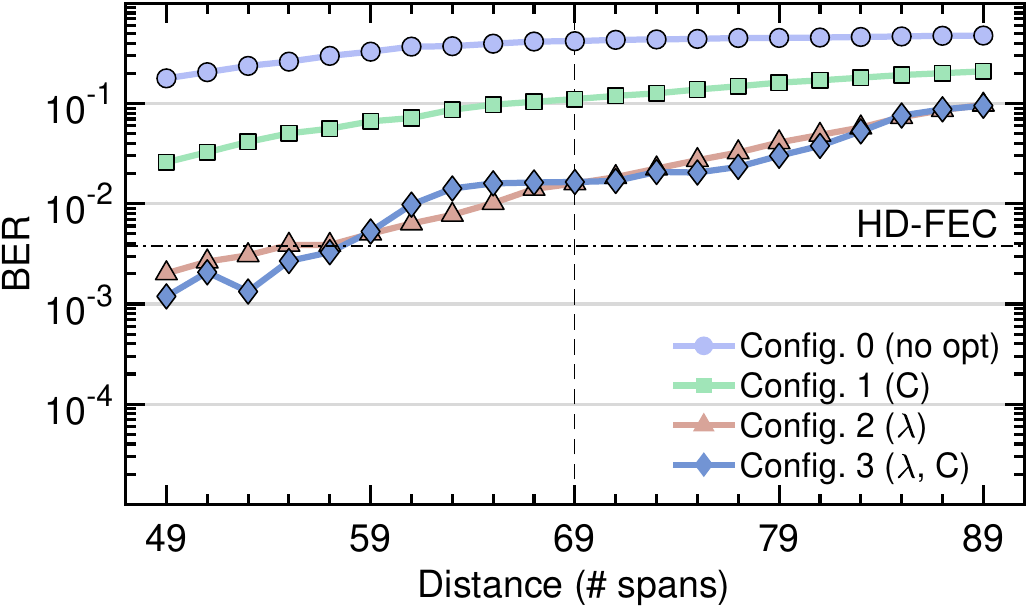}
    }
    \subfloat[Power factor optimized ($\nonlinfactlpa =
\hat{\gamma}_{E2E}$)]{\label{fig:ber_vs_distance_true}
        \includegraphics[width=.49\linewidth]{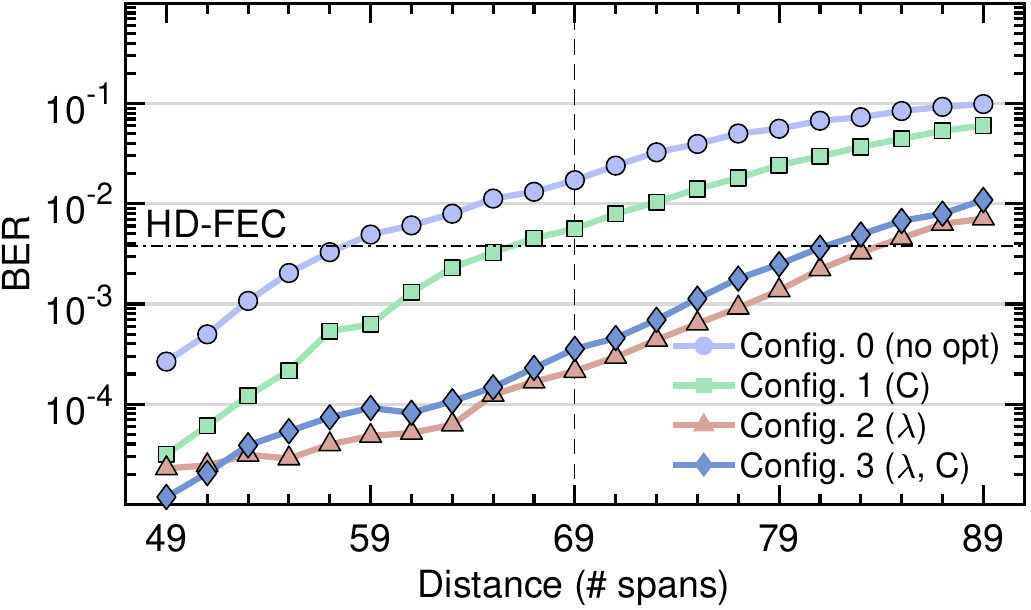}
    }
    \caption{\emph{Test performance \ac{NN} receiver} - \ac{BER} as a function of the transmission distance. The vertical dashed line marks the optimization distance. }
    \label{fig:ber_vs_distance}
\end{figure*}

\subsection{\Acl{NN} receiver \ac{BER} performance}
\label{subsec:results_nn}

The performances of the communication system, which was trained at a distance of $69 \times 80$~km, have  been evaluated over a range of transmission distances from $49 \times 80$~km to $89 \times 80$~km in order to verify the robustness of the optimized transmitter to a transmission distance mismatch.
The transmitter of the system uses the parameters optimized by the \ac{AE} (eigenvalues, $\nftb[i]$-constellations, and $\nonlinfactlpa$).
% The nonlinear channel was simulated using the \ac{SSFM} with adaptive step size that guarantees a maximum nonlinear phase rotation of up to 0.01 degrees.
The receiver \ac{NN} used for the performance evaluation  has the same topology as the one described in Section~\ref{sec:nn_receiver} but it was re-trained independently for each of the transmission distances considered. The re-training was performed using a training dataset of $500 \times 10^3$ symbols,
%a cross-validation dataset of $100 \times 10^3$ symbols,
a batch size of 1000 symbols, and 1000 training iterations. The re-training is necessary for the different transmission lengths\ul{, as the different amounts of accumulated dispersion and nonlinear phase shift lead to significant variations in the time-domain waveforms seen by the receiver. The re- training} was also performed for $69 \times 80$~km.
As already shown in \cite{Karanov2018}, re-training the receiver \ac{NN} once the transmitter parameters have been optimized by the \ac{AE} improves the performance. An intuitive explanation is that during the training of the \ac{AE}, both the transmitter and the receiver \ac{NN} parameters are changed at each training iteration so that the final \ac{NN} is effectively trained for the specific transmitter parameters only for a single batch. The small batch size may not allow converging to the optimal set of weights. The re-training is particularly useful in this work given the very limited size of the batch used during the training of the \ac{AE}. \ul{Larger batch sizes may provide improved performance potentially removing the need for re-training at  $69 \times 80$~km~}\cite{OShea2018}. 
\ul{Note that, alternatively to re-training for each transmission distance, the receiver }\ac{NN} \ul{could be trained with a dataset containing samples from several transmission distances. As shown in~}\cite{Karanov2018}\ul{, this may lead to stronger robustness to variation in the target link length, but overall sub-optimal performance compared to selectively training for each distance independently.}
After re-training, the \ac{BER} has been computed using a testing dataset of $5 \cross 10^5$ symbols.

\figurename~\ref{fig:ber_vs_distance} shows the \ac{BER} performance  as a function of the transmission distance for \subref{fig:ber_vs_distance_false} the transmitter power un-optimized scenario
($\nonlinfactlpa~=~\nonlinfactlpa[LPA]$) and \subref{fig:ber_vs_distance_true} the power-optimized scenario ($\nonlinfactlpa~=~\nonlinfactlpa[E2E]$).

In \figurename~\ref{fig:ber_vs_distance}~\subref{fig:ber_vs_distance_false} ($\nonlinfactlpa~=~\nonlinfactlpa[LPA]$), the worst performance is shown by \cfg{0}, i.e. when none of the transmitter parameters are optimized. The \ac{BER} is close to 0.5 at the trained distance and drops only slightly for shorter transmission distances. When the $\nftb[i]$-constellations are optimized (\cfg{1}), the performance is improved compared to \cfg{0}, but  the \ac{BER} is still above $1 \times 10^{-2}$ over the full range of distances considered. Separating the solitonic components in time provides only a slight improvement.
When the \ac{AE} optimizes the imaginary part of the eigenvalues (\cfg{2}), the
\ac{BER} at optimization distance is further reduced by almost one order of magnitude.
Finally when both the eigenvalues and the $\nftb[i]$-constellations  are optimized (\cfg{3}) the \ac{BER} is similar to that \cfg{2} for all the distances considered, consistently with the training performance.
The optimization of the imaginary part of the eigenvalue, which controls the transmitted power, plays a key role on the system performance.
The fact that a
reduction of the signal power reduces the \ac{BER} hints that the system performance
at the optimization distance might be more limited by nonlinear effects and not by the
 \ac{OSNR} so that reducing the transmitted power improves the performance. % Follow-up?

In \figurename~\ref{fig:ber_vs_distance}~\subref{fig:ber_vs_distance_true}
the curves for the second scenario ($\nonlinfactlpa~=~\nonlinfactlpa[E2E]$) are shown. At the optimization distance,
all configurations show an improved \ac{BER}, with values of  $1.72 \cross 10^{-2}$, $5.63 \cross 10^{-3}$, $2.14 \cross 10^{-4}$ and $3.56 \cross 10^{-4}$ for \emph{configurations} 0 to 3, respectively.
The relative performance between the four configurations follows  the discussion reported for the $\nonlinfactlpa = \nonlinfactlpa[LPA]$ scenario, but the additional degree of freedom provides a significant improvement for all the configurations.
This proves that the average launch power (through the parameter $\nonlinfactlpa$) is a critical optimization parameter. The \ac{BER} obtained in the best case (\cfg{2}, $\nonlinfactlpa = \nonlinfactlpa[E2E]$) is three orders of
magnitudes lower than the reference case (\cfg{0}, $\nonlinfactlpa = \nonlinfactlpa[LPA]$) where none of the transmitter parameters are optimized. This best configuration among those considered allows reaching a transmission distance more than three times what is achievable with a \ul{manually-optimized} \ac{NFT} receiver (approx. 83 spans vs. the 25 spans of  \figurename{}~\ref{fig:ber_vs_distance_nft}).

We can observe in \figurename~\ref{fig:ber_vs_distance}~\subref{fig:ber_vs_distance_true} that the \cfg{3} performs slightly worse than the \cfg{2}, despite having
more optimization degrees of freedom. This is likely due to the presence of local optima in the cost function from where the optimization procedure was not able to escape. This result further justifies our choice of guiding the encoder rather than performing a fully blind optimization. \ul{The re-training of the receiver can only partially improve on the negative impact of local optima as the transmitter parameters are not improved and they have a clear impact on the overall performance.} In the future, strategies to move towards this latter goal without suffering from local optima in the optimization landscape need to be devised \cite{li2018visualizing}. %using \iac{NFT}-based transmitter (theoretically ideal for a lossless/noiseless channel) to guide the end-to-end optimization.
\ul{Overall, the comparison between }\cfg{1} \ul{and } \cfg{2}, \ul{under both optimization conditions ($\nonlinfactlpa = \nonlinfactlpa[LPA]$ and $\nonlinfactlpa = \nonlinfactlpa[E2E]$), hints that the optimization of the eigenvalues is more critical than the optimization of the spectral amplitudes.}

For all the eight different configurations tested we can observe that the performance
gain is preserved across all the transmission distances between $49$ and  $89 \cross
80$~km, even though the optimization was performed at 69 spans, showing the robustness
of the transmitter parameters optimization to the transmission distance.

% \section{Discussion}
% \label{sec:discussion}
% (solitons are no solitons any more)
% This block based receiver does not account for memory between adjacent symbols that is
% possibly introduced by the channel. Nonetheless this limit should automatically be
% taken into account by the end-to-end optimizer that will be forced to find particular
% solutions of parameters that allow the signal to not disperse.

% Gradient averaging can be used to fine tune the performance. Actually it is implemented!
% \subsection{Bandwidth constraints}

% Talk about the fact that the optimal constellations have lower power. No bandwidth
% constraint.

% The constellations should have been optimized
% using a NFT receiver, though the problem of the NFT receiver is that I do not know how to make
% it differentiable (e.g. eigenvalue search).

\section{Conclusion}
\label{sec:conclusions}

In this work, we proposed for the first time an \ac{AE} scheme for coherent fiber-optic communications considering the accurate model of a nonlinear dispersive fiber channel. The system uses \iac{NFDM} transmitter that performs the symbol-to-waveform encoding and an \ac{NN}-based receiver that performs the waveform-to-symbol decoding. The chosen transmitter constraints the solutions space of the generated time-domain waveforms to solitonic pulse-shapes, facilitating the optimization of the \ac{AE}. Moreover the minimal dispersion of the solitons, even in the presence of losses, allows to use a memory-less receiver implemented with a low-complexity \ac{NN}. The full optical nonlinear channel model has been  implemented using the \ac{SSFM}. Given that this method is a composition of basic differentiable operations, it was possible to use the automatic differentiation capabilities of the Tensorflow library to perform the training of the \ac{AE} across this channel model.

The \ac{AE} has been trained using $2~\times $ four different configurations of the transmitter trainable parameters (eigenvalues, $\nftb[i]$-constellations, and $\nonlinfactlpa$). The numerical results of the performance testing of the system demonstrated that the best configuration of these parameters allows a reduction of three order of magnitude in \ac{BER} at the optimization distance compared to the un-optimized transmitter configuration. In particular, it was shown that the system performance is particularly sensitive to the imaginary part of the eigenvalues.
Moreover, it was shown that the best results are obtained when the \ac{AE} is also free to optimize the waveform launch power through the optimization of $\nonlinfactlpa$.
Compared to \ul{a manually-optimized} \ac{NFDM} communication system used as a benchmark, the proposed proof-of-concept system allows extending the transmission reach from 2000 to 6640~km at the \ac{HD-FEC} threshold.

We believe that this work moves the research on the end-to-end optimization of communication systems a step closer to the final goal of realizing a general
\ac{AE} communication scheme employing \acp{NN} for both the transmitter and the receiver and defined over a nonlinear dispersive fiber-optic channel.

\section*{Acknowledgment}
This work is supported by the European Research Council through the ERC-CoG FRECOM
project (grant agreement no. 771878) and by the Villum Foundation through the Villum Young Investigator fellowship OPTIC-AI (grant no. 29344).

\ifCLASSOPTIONcaptionsoff
  \newpage
\fi

%% CUSTOM BIBLIOGRAPHY
\printbibliography
%% END CUSTOM BIBLIOGRAPHY

\end{document}

%% file: acronyms.tex
\usepackage{acro}

\DeclareAcronym{AD}{short=AD,long=automatic differentiation,short-indefinite=an}
\DeclareAcronym{ADC}{short=ADC,long=analog-to-digital converter,short-indefinite=an}
\DeclareAcronym{ADM}{short=ADM,long=add-drop multiplexer,short-indefinite=an}
\DeclareAcronym{AE}{short=AE,long=autoencoder,short-indefinite=an}
\DeclareAcronym{AOM}{short=AOM,long=acusto-optic modulator,short-indefinite=an}
\DeclareAcronym{ASE}{short=ASE,long=amplified spontaneous emission,short-indefinite=an}
\DeclareAcronym{AIR}{short=AIR,long=achievable information rate,short-indefinite=an}
\DeclareAcronym{AKNS}{short=AKNS,long=Ablowitz-Kaup-Newell-Segur,short-indefinite=an}
\DeclareAcronym{AWG}{short=AWG,long=arbitrary waveform generator,short-indefinite=an}
\DeclareAcronym{AWGN}{short=AWGN,long=additive white Gaussian noise,short-indefinite=an}
\DeclareAcronym{B2B}{short=B2B,long=back-to-back}
\DeclareAcronym{BER}{short=BER,long=bit error ratio}
\DeclareAcronym{BPD}{short=BPD,long=balanced photodetector}
\DeclareAcronym{BPS}{short=BPS,long=blind phase search}
\DeclareAcronym{CFO}{short=CFO,long=carrier frequency offset}
\DeclareAcronym{DAC}{short=DAC,long=digital-to-analog converter}
\DeclareAcronym{DC}{short=DC,long=direct current}
\DeclareAcronym{DCF}{short=DCF,long=dispersion-compensating fiber}
\DeclareAcronym{DBP}{short=DBP,long=digital back-propagation}
\DeclareAcronym{DFT}{short=DFT,long=discrete Fourier transform}
\DeclareAcronym{DP-NFDM}{short=DP-NFDM,long=dual-polarization nonlinear frequency division multiplexing}
\DeclareAcronym{DSO}{short=DSO,long=digital storage oscilloscope}
\DeclareAcronym{DSP}{short=DSP,long=digital signal processing}
\DeclareAcronym{DT}{short=DT,long=Darboux transform}
\DeclareAcronym{E2E}{short=E2E,long=end-to-end,short-indefinite=an}
\DeclareAcronym{EDC}{short=EDC,long=electrical dispersion compensation,short-indefinite=an}
\DeclareAcronym{EDFA}{short=EDFA,long=erbium-doped fiber amplifier,short-indefinite=an}
\DeclareAcronym{ENOB}{short=ENOB,long=effective number of bits,short-indefinite=an}
\DeclareAcronym{EVM}{short=EVM,long=error vector magnitude,short-indefinite=an}
\DeclareAcronym{FEC}{short=FEC,long=forward error correction,short-indefinite=an}
\DeclareAcronym{FFT}{short=FFT,long=fast Fourier transform,short-indefinite=an}
\DeclareAcronym{FL}{short=FL,long=fiber laser,short-indefinite=an}
\DeclareAcronym{FWHM}{short=FWHM,long=full-width half-maximum,short-indefinite=an}
\DeclareAcronym{FWM}{short=FWM,long=four-wave mixing,short-indefinite=an}
\DeclareAcronym{GLM}{short=GLM,long=Gelfand-Levitan-Marchenko}
\DeclareAcronym{GLME}{short=GLME,long=Gelfand-Levitan-Marchenko equation}
\DeclareAcronym{GPU}{short=GPU,long=graphics processing unit}
\DeclareAcronym{GVD}{short=GVD,long=group velocity dispersion}
\DeclareAcronym{HD-FEC}{short=HD-FEC,long=hard-decision forward error correction,short-indefinite=an}
\DeclareAcronym{I}{short=I,long=in-phase,short-indefinite=an}
\DeclareAcronym{IFFT}{short=IFFT,long=inverse fast Fourier transform,short-indefinite=an}
\DeclareAcronym{IM}{short=IM,long=intensity modulation,short-indefinite=an}
\DeclareAcronym{INFT}{short=INFT,long=inverse nonlinear Fourier transform,short-indefinite=an}
\DeclareAcronym{ISI}{short=ISI,long=inter-symbol interference,short-indefinite=an}
\DeclareAcronym{ISO}{short=ISO,long=isolator}
\DeclareAcronym{IST}{short=IST,long=inverse scattering transform,short-indefinite=an}
\DeclareAcronym{IVP}{short=IVP,long=initial value problem,short-indefinite=an}
\DeclareAcronym{LO}{short=LO,long=local oscillator,short-indefinite=an}
\DeclareAcronym{LMMSE}{short=LMMSE,long=linear minimum mean square error,short-indefinite=an}
\DeclareAcronym{LPA}{short=LPA,long=lossless path-averaged,short-indefinite=an}
\DeclareAcronym{ME}{short=ME,long=Manakov equation,short-indefinite=an}
\DeclareAcronym{MS}{short=MS,long=Manakov system,short-indefinite=an}
\DeclareAcronym{MAP}{short=MAP,long=maximum a posteriori,short-indefinite=a}
\DeclareAcronym{ML}{short=ML,long=maximum likelihood,short-indefinite=an}
\DeclareAcronym{MZM}{short=MZM,long=Mach-Zehnder modulator,short-indefinite=an}
\DeclareAcronym{MZSP}{short=MZSP,long=Manakov-Zakharov-Shabat spectral problem,short-indefinite=an}
\DeclareAcronym{NFDM}{short=NFDM,long=nonlinear frequency division multiplexing,short-indefinite=an}
\DeclareAcronym{NFT}{short=NFT,long=nonlinear Fourier transform,short-indefinite=an}
\DeclareAcronym{NCG}{short=NCG,long=Nystrom conjugate gradient,short-indefinite=an}
\DeclareAcronym{NIS}{short=NIS,long=nonlinear inverse synthesis,short-indefinite=an}
\DeclareAcronym{NLSE}{short=NLSE,long=nonlinear Schr\"odinger equation,short-indefinite=an}
\DeclareAcronym{NMSE}{short=NMSE,long=normalized mean squared error,short-indefinite=an}
\DeclareAcronym{NN}{short=NN,long=neural network,short-indefinite=an}
\DeclareAcronym{OBPF}{short=OBPF,long=optical band pass filter,short-indefinite=an}
\DeclareAcronym{ODE}{short=ODE,long=ordinary differential equation,short-indefinite=an}
\DeclareAcronym{OFDM}{short=OFDM,long=orthogonal frequency-division multiplexing,short-indefinite=an}
\DeclareAcronym{OPC}{short=OPC,long=optical phase conjugation,short-indefinite=an}
\DeclareAcronym{OSNR}{short=OSNR,long=optical signal-to-noise ratio,short-indefinite=an}
\DeclareAcronym{PAPR}{short=PAPR,long=peak-to-average power ratio}
\DeclareAcronym{PC}{short=PC,long=polarization controller}
\DeclareAcronym{PDE}{short=PDE,long=partial differential equation}
\DeclareAcronym{PMD}{short=PMD,long=polarization mode dispersion}
\DeclareAcronym{ppm}{short=ppm,long=parts-per-million}
\DeclareAcronym{PRBS}{short=PRBS,long=pseudo-random bit sequence}
\DeclareAcronym{PSK}{short=PSK,long=phase-shift keying}
\DeclareAcronym{Q}{short=Q,long=quadrature}
\DeclareAcronym{QAM}{short=QAM,long=quadrature amplitude modulation}
\DeclareAcronym{QPSK}{short=QPSK,long=quadrature phase-shift keying}
\DeclareAcronym{SDM}{short=SDM,long=space-division multiplexing,short-indefinite=an}
\DeclareAcronym{SELU}{short=SELU,long=scaled exponential linear units,short-indefinite=an}
\DeclareAcronym{SSMF}{short=SSMF,long=standard single-mode fiber,short-indefinite=an}
\DeclareAcronym{SMF}{short=SMF,long=single-mode fiber,short-indefinite=an}
\DeclareAcronym{SNR}{short=SNR,long=signal-to-noise ratio,short-indefinite=an}
\DeclareAcronym{S/P}{short=S/P,long=serial-to-parallel,short-indefinite=an}
\DeclareAcronym{SPM}{short=SPM,long=self-phase modulation,short-indefinite=an}
\DeclareAcronym{SSFM}{short=SSFM,long=split-step Fourier method,short-indefinite=an}
\DeclareAcronym{WDM}{short=WDM,long=wavelength-division multiplexing}
\DeclareAcronym{XPM}{short=XPM,long=cross-phase modulation,short-indefinite=an}
\DeclareAcronym{ZSP}{short=ZSP,long=Zakharov-Shabat spectral problem}
\DeclareAcronym{Z-S}{short=Z-S,long=Zakharov-Shabat}

%% file: commands-optcomm.tex
\usepackage{xargs}        % To define commands with more than one optional value \newcommandx
\usepackage{etoolbox}     % (Modern package) For \ifstrequal
\usepackage{physics}      % For \Re \Im as Re() Im() and tons of operators,
\usepackage{amsfonts}     % \mathbb

\newcommand{\gbaud}{GBd}

\newcommand{\iunit}{j}                      % Imaginary unit
\newcommand{\vecbold}[1]{                   % Vector / Matrix bold notation. Remove \boldsymbol
  {{#1}}}                        % for single pol papers

\newcommand{\loss}{\alpha}                  % Loss coefficient (alpha)
\newcommand{\dispersion}{\beta_2}           % Group velocity dispersion
\newcommand{\nonlinfact}{\gamma}            % Nonlinear coefficient (gamma)
\newcommand{\lspan}{L_{s}}                  % Span length
\newcommand{\nspans}{N_s}                   % Number of spans

\newcommand{\ttm}{\tau}                        % Time
\newcommand{\ssp}{\ell}                        % Space
\newcommandx{\flds}[2][1, 2]{                  % Electrical field (short) [1 = Mode index, 2 = Quantity modifier]
    \des[#2]{E}{#1}}
\newcommandx{\fldsrc}[3][1 ,2, 3]{             % Electrical field/signal, don't use it explicitly
  \flds[#1][#3](\ttm_{#2}}                     % [1 = Mode index, 2 = Time variable index, 3 = Quantity modifier]
\newcommandx{\fld}[3][1, 2, 3]{                % Electrical field/signal
\fldsrc[#1][#2][#3] )}                         % [1 = Mode index, 2 = Time variable index, 3 = Quantity modifier]
\newcommandx{\fldl}[3][1, 2, 3]{               % Electrical field/signal (long)
\fldsrc[#1][#2][#3], \ssp) }                   % [1 = Mode index, 2 = Time variable index, 3 = Quantity modifier]

\newcommand\des[3][]{\desaux{#2}{#3}#2_\relax{#1}}
\def\desaux#1#2#3_#4\relax#5{%
  \ifx\relax#4\relax
    \ifx\relax#2\relax
      \vecbold{#5#1}
    \else
      #5#1_{#2}
    \fi
  \else
    \ifx\relax#2\relax
      \vecbold{#5#3_{\stripus#4}}
    \else
      #5#3_{\stripus#4,#2}
    \fi
  \fi
}
\def\stripus#1_{#1}

%% file: commands-nft.tex
\DeclareMathOperator*{\inft}{INFT}          % INFT

\newcommandx{\nonlinfactlpa}[1][1]{\bar{\gamma}_{#1}}   % LPA equivalent nonlinear coefficient (gamma)

\newcommand{\nttm}{t}                       % Normalized time
\newcommand{\nssp}{z}                       % Normalized space
\newcommand{\To}{T_{0}}                     % NFT normalization time
\newcommand{\nftL}{\mathcal{L}}             % NFT normalization length

\newcommand{\nfldletter}{q}
\newcommandx{\nflds}[2][1, 2]{              % Normalized electrical field (short) [1 = Mode index, 2 = Quantity modifier]
    \des[#2]{\nfldletter}{#1}}
\newcommandx{\nfldsrc}[3][1 ,2, 3]{         % Normalized electrical field/signal, don't use it explicitly
  \nflds[#1][#3](\nttm_{#2}}                % [1 = Mode index, 2 = Time variable index, 3 = Quantity modifier]
\newcommandx{\nfld}[3][1, 2, 3]{            % Normalized electrical field/signal
  \nfldsrc[#1][#2][#3] )}                   % [1 = Mode index, Time variable index, 3 = Quantity modifier]
\newcommandx{\nfldl}[3][1, 2, 3]{           % Normalized electrical field/signal (long)
  \nfldsrc[#1][#2][#3], \nssp )}            % [1 = Mode index, Time variable index, 3 = Quantity modifier]

\newcommandx{\nftquantity}[5][1, 2, 3, 4, 5]{       % NFT quantity, don't use it explicitly
  \des[#4]{#1}{#3}(\eig[#2][#5]}                    % [1 = Quantity, 2 = eigenvalue index,
\newcommandx{\eig}[2][1, 2]{#2{\lambda}_{#1}}  % Nonlinear frequency / Eigenvalue [1 = eigenvalue index]

\newcommand{\nftasrc}{a}                       % NFT coefficient 'a', don't use it explicitly
\newcommandx{\nfta}[2][1, 2]{#2{\nftasrc}(\eig[#1])}         % NFT coefficient 'a' [1 = eigenvalue index, 2 = Quantity modifier]
\newcommandx{\nftaderiv}[2][1, 2]{#2{\nftasrc}'(\eig[#1])}   % NFT coefficient 'a1' [1 = eigenvalue index, 2 = Quantity modifier]

\newcommandx{\nftbsrc}[3][1= ,2=, 3=]{       % NFT coefficient 'a', don't use it explicitly
  \nftquantity[b][#1][#2][#3]}
\newcommandx{\nftb}[3][1= ,2=, 3=]{          % NFT b scattering coeff. [1 = eigen index, 2 = mode index, 3 = modifier]
  \nftbsrc[#1][#2][#3] ) }
\newcommandx{\nftbl}[3][1= ,2=, 3=]{         % NFT b scattering coeff. [1 = eigen index, 2 = mode index, 3 = modifier]
  \nftbsrc[#1][#2][#3], \nssp) }

\newcommandx{\cnftsrc}[2][1= ,2=]{          % NFT continuous spectrum, don't use it explicitly
  \nftquantity[Q_c][][#1][#2] }
\newcommandx{\cnft}[2][1= ,2=]{             % NFT cont. spectrum [1 = mode index, 2 = modifier]
  \cnftsrc[#1][#2] ) }
\newcommandx{\cnftl}[2][1= ,2=]{            % NFT continuous spectrum, long version [1 = mode index, 2 = modifier]
  \cnftsrc[#1][#2], \nssp) }

\newcommandx{\dnftsrc}[3][1=, 2=, 3=]{      % NFT discrete spectrum, don't use it explicitly
  \nftquantity[Q_d][#1][#2][#3] }
\newcommandx{\dnft}[3][1=, 2=, 3=]{         % NFT disc. spectrum [1 = eigen index, 2 = mode index, 3 = modifier]
  \dnftsrc[#1][#2][#3] ) }
\newcommandx{\dnftl}[3][1=, 2=, 3=]{        % NFT disc. spectrum long [1 = eigen index, 2 = mode index, 3 = modifier]
  \dnftsrc[#1][#2][#3], \nssp) }

\newcommandx{\nftInvH}[2][1=, 2=]{e^{4\iunit\eig[#1]^2#2}}   % Inverse NFT spectrum space evolution (z = 1) / Inverse channel transfer function
\newcommandx{\nftInvHtext}[2][1=, 2=]{\exp(4\iunit\eig[#1]^2#2)}  % Inverse NFT spectrum space evolution (z = 1) / Inverse channel transfer function (text version)

\newcommandx{\nftbconj}[2][1=, 2=]{        % NFT coefficient 'b' conjugate [1 = eigenvalue index, 2 = mode index]
\nftb[#1][#2][\bar]}

\newcommand{\estoperator}{\widehat}        % Estimation operator

\newcommandx{\nftbest}[2][1=, 2=]{         % NFT b scattering coeff. [1 = eigen index, 2 = mode index]
    \nftbsrc[#1][#2]
    [\estoperator][\estoperator] ) }
\newcommandx{\cnftest}[1][1=]{             % NFT cont. spectrum [1 = mode index]
    \cnftsrc[#1]
    [\estoperator][\estoperator] ) }
\newcommandx{\dnftest}[2][1=, 2=]{         % NFT disc. spectrum long [1 = eigen index, 2 = mode index]
    \dnftsrc[#1][#2]
    [\estoperator][\estoperator] ) }

\newcommandx{\eigf}[2][1, 2]{\des[#2]{v}{#1}} % Eigenfunction [1 = Component index, 2 = modifier ]

\newcommandx{\jostquantity}[3][1, 2, 3]{    % Jost quantity, don't use it explicitly
  \des[#3]{#1}{#2}}                         % [1 = Quantity, 2 = component index,
\newcommand{\jostconjoperator}{\bar}            % Jost 'conjugation' operator

\newcommandx{\jostps}[2][1 ,2]{             % Jost sol. P(ositive) short [1 = component index, 2 = modifier]
  \jostquantity[\psi][#1][#2]}
\newcommandx{\jostpsrc}[2][1 ,2]{           % Jost sol. P, don't use it explicitly [1 = component index, 2 = modifier]
  \jostps[#1][#2](\nttm}
\newcommandx{\jostp}[2][1 ,2]{              % Jost sol. P [1 = component index, 2 = modifier]
  \jostpsrc[#1][#2] ) }
\newcommandx{\jostpl}[3][1 ,2, 3]{          % Jost sol. P with space [1 = component index, 2 = eigen index, 3 = modifier]
  \jostpsrc[#1][#3], \eig[#2] ) }
\newcommandx{\jostpll}[3][1 ,2, 3]{         % Jost sol. P with space and field [1 = component index, 2 = eigen index, 3 = modifier]
  \jostpsrc[#1][#3], \eig[#2]; \nflds ) }

\newcommandx{\jostpconjs}[2][1, 2]{         % Jost sol. P conjugate (short) [1 = component index, 2 = eigen index, 3 = modifier]
\jostps[#1][\jostconjoperator]}
\newcommandx{\jostpconj}[2][1, 2]{          % Jost sol. P conjugate [1 = component index, 2 = eigen index, 3 = modifier]
\jostp[#1][\jostconjoperator]}
\newcommandx{\jostpconjl}[2][1, 2]{         % Jost sol. P conjugate (long) [1 = component index, 2 = eigen index]
\jostpl[#1][#2][\jostconjoperator]}

\newcommandx{\jostns}[2][1 ,2]{             % Jost sol. N(egative) (short) [1 = component index, 2 = modifier]
  \jostquantity[\phi][#1][#2]}
\newcommandx{\jostnsrc}[2][1 ,2]{           % Jost sol. N, don't use it explicitly [1 = component index, 2 = modifier]
  \jostns[#1][#2](\nttm}
\newcommandx{\jostn}[2][1 ,2]{              % Jost sol. N [1 = component index, 2 = modifier]
  \jostnsrc[#1][#2] ) }
\newcommandx{\jostnl}[3][1 ,2, 3]{         % Jost sol. N with space [1 = component index, 2 = eigen index, 3 = modifier]
  \jostnsrc[#1][#3], \eig[#2] ) }
\newcommandx{\jostnll}[3][1 ,2, 3]{        % Jost sol. N with space and field [1 = component index, 2 = eigen index, 3 = modifier]
  \jostnsrc[#1][#3], \eig[#2]; \nflds ) }

\newcommandx{\jostnconjs}[2][1, 2]{          % Jost sol. N conjugate (short) [1 = component index, 2 = eigen index, 3 = modifier]
\jostns[#1][\jostconjoperator]}
\newcommandx{\jostnconj}[2][1, 2]{          % Jost sol. N conjugate [1 = component index, 2 = eigen index, 3 = modifier]
\jostn[#1][\jostconjoperator]}
\newcommandx{\jostnconjl}[2][1, 2]{         % Jost sol. N conjugate (long) [1 = component index, 2 = eigen index]
\jostnl[#1][#2][\jostconjoperator]}

\newcommand{\darbauxsolsrc}{\nu}
\newcommand{\darboperator}{\widehat}           % Darboux operator
\newcommandx{\darbauxsols}[3][1, 2, 3]{        % Darboux transform auxiliary solution
  \des[#3]{\darbauxsolsrc}{#1}^{#2}                       % [1 = Component index, 2 = Superscript, 3 = Modifier]
}
\newcommandx{\darbauxsolp}[3][1, 2, 3]{        % Darboux transform auxiliary solution boundary plus inf
  \darbauxsols[#1][+][#3](                    % [1 = Component index, 2 = Eigen index, 3 = Modifier]
    \eig[#2]
  )}
\newcommandx{\darbauxsoln}[3][1, 2, 3]{        % Darboux transform auxiliary solution boundary minus inf
  \darbauxsols[#1][-][#3](                    % [1 = Component index, 2 = Eigen index, 3 = Modifier]
    \eig[#2]
  )}
\newcommandx{\darbauxsol}[4][1, 2, 3, 4]{      % Darboux transform auxiliary solution
  \darbauxsols[#1][#2][#4](                    % [1 = Component index, 2 = Superscript, 3 = Time variable, 4 = Modifier]
    \ifstrequal{#3}{}{\nttm}{#3}
  )}
\newcommandx{\darbauxsoll}[5][1, 2, 3, 4, 5]{  % Darboux transform auxiliary solution
  \darbauxsols[#1][#3][#5](                    % [1 = Component index, 2 = eigen index,
    \ifstrequal{#4}{}{\nttm}{#4},              %  3 = Superscript, 4 = Time variable, 5 = Modifier]
  \eig[#2])}
\newcommandx{\nflddarbs}[1][1]{         % Normalized field/signal modified by darboux [1 = Mode index, 2 = Time variable index ]
  \nflds[#1][\darboperator] }
\newcommandx{\nflddarb}[2][1, 2]{       % Normalized field/signal modified by darboux [1 = Mode index, 2 = Time variable index ]
  \nfld[#1][#2][\darboperator] }
\newcommandx{\eigfdarb}[1][1]{          % Eigenfunction modified by darboux [1 = Component index ]
\eigf[#1][\darboperator] }
\newcommand{\predarboperator}{\widetilde}

\newcommandx{\cnftpredarb}[1][1=]{      % NFT continuous spectrum pre-distrted for Darboux [1 = mode index]
  \cnft[#1][\predarboperator] }
\newcommandx{\nfldpredarbs}[2][1, 2]{   % Normalized electrical field/signal pre-distrted for Darboux (short)
\nflds[#1][\predarboperator]}      % [1 = mode index]
\newcommandx{\nfldpredarb}[2][1, 2]{    % Normalized electrical field/signal pre-distrted for Darboux
\nfld[#1][#2][\predarboperator]}       % [1 = mode index, 2 = Time variable index]
\newcommandx{\nfldpredarbl}[2][1, 2]{   % Normalized electrical field/signal pre-distrted for Darboux (long)
\nfldl[#1][#2][\predarboperator]}      % [1 = mode index, 2 = Time variable index]
\newcommandx{\jostnpredarbll}[3][1 ,2, 3]{        % Jost sol. N with space and field [1 = component index, 2 = eigen index, 3 = modifier]
  \jostnsrc[#1][#3], \eig[#2]; \nflds[][\predarboperator] ) }
\newcommandx{\jostppredarbll}[3][1 ,2, 3]{        % Jost sol. P with space and field [1 = component index, 2 = eigen index, 3 = modifier]
  \jostpsrc[#1][#3], \eig[#2]; \nflds[][\predarboperator] ) }

\newcommand{\cradius}[1][]{r_{#1}}               % b-constellation radius
\newcommand{\cphase}[1][]{\phi_{#1}}               % b-constellation phase
\newcommand{\cphasediff}{\Delta \cphase}               % b-constellation phase difference

%% file: commands-ml.tex
\newcommand{\lr}{\eta}

\newcommand{\symbalphabet}{X}                % Symbol alphabet
\newcommand{\txsymbonehot}{x}       % Tx symbol one-hot-encoded
\newcommand{\rxsymbonehot}{\hat{x}} % Rx symbol one-hot-encoded
\newcommand{\rxsymbwave}{\mathbf{y}}         % Rx symbol waveform

\DeclareMathOperator*{\xentropy}{xentr}          % Cross-entropy
\DeclareMathOperator*{\nadam}{Nadam}             % Nesterov Adam optimizer

\newcommand{\trainparams}{\mathbf{\theta}}
\newcommand{\nntrainparams}{\mathbf{w}_{NN}}
\newcommand{\ntrainiter}{N_{iter}}

%% file: algorithms/ae_training.tex
%!TEX root = ../main.tex
\begin{algorithm}[t]
    \caption{End-to-end training procedure}
    \hspace*{\algorithmicindent} \ul{\textbf{Inputs:} Trainable parameters  $\trainparams^{(0)}$}\\ % CORRECT? @ SIMGA
    \hspace*{\algorithmicindent} \ul{\textbf{Outputs:} Optimized trainable   parameters $\trainparams^{(\ntrainiter)}$} % CORRECT? @ SIMGA 
    \begin{algorithmic}[1]
        \State Initialize trainable parameters
        \State $[C_i, \eig[i]]\, i=1,2 \gets$ configuration 0 in \tablename{}~\ref{tab:configurations}
        \State $\nonlinfactlpa \gets \nonlinfact$
        \State $\nntrainparams \gets$ \emph{Glorot} uniform initialization
        \State Initialize feature vector $\trainparams^{(0)}$ as in Table~\ref{tab:configurations}
        \For{$n \gets 1,\,\dots,\,\ntrainiter$} 
        \State Generate B random symbols $\{\txsymbonehot\}_{B}$ % B = 64 in text
        \State Map $\{\txsymbonehot\}_{B}$ to $\{\nftb[i]\}_{B}$ using  $C_i^{(n)}, i=1,2$
        \State Generate B waveforms $\fld=\inft(\eig[i], \nftb[i])$ 
        \State Construct channel input $s(\ttm)$ by serializing $\{\fld\}_B$
        \State Propagate $s(\ttm)$ in the channel using \ac{SSFM} to get \ul{$r(\ttm)$}
        \State Slice $r(\ttm)$ into B waveforms $\{\rxsymbwave\}_{B}$
        \State Detect $\{\rxsymbwave\}_{B}$ using rx NN with $\nntrainparams$ 
        \State Rx NN outputs B symbols $\{\rxsymbonehot\}_{B}$
        \State Compute cross-entropy $L(\theta)$ = $\xentropy(\{\txsymbonehot\}_{B}, \{\rxsymbonehot\}_{B})$
        \State Backpropagate to compute gradient $\gradient_\trainparams \tilde{L}(\trainparams^{(n)})$
        \State Compute $\trainparams^{(n+1)} \gets \trainparams^{(n)} - \eta\,\nadam( \gradient_{\trainparams} \tilde{L}(\trainparams^{(n)}))$
        \State Update trainable parameters from $\trainparams^{(n+1)}$
        \EndFor
    \end{algorithmic}
    \label{alg:ae_training}
\end{algorithm}

%% file: bibliography-gscholar.bib
@book{goodfellow2016deep,
  title={Deep learning},
  author={Goodfellow, Ian and Bengio, Yoshua and Courville, Aaron},
  year={2016},
  publisher={MIT press}
}

@book{hasegawa1995solitons,
  title={Solitons in optical communications},
  author={Hasegawa, Akira and Kodama, Y{\=u}ji},
  number={7},
  year={1995},
  publisher={Oxford University Press, USA}
}

@book{AgrawalBook,
  title={Nonlinear Fiber Optics},
  author={Agrawal, G P},
  edition=5,
  year={2013},
  publisher={Academic Press}
}

@book{matveev1991darboux,
  title={Darboux transformations and solitons},
  author={Matveev, Vladimir B and Salle, M A},
  year={1991},
  publisher={Springer-Verlag}
}

@article{MetodiJLT16,
author={M. P. {Yankov} and F. {Da Ros} and E. P. {da Silva} and S. {Forchhammer} and K. J. {Larsen} and L. K. {Oxenl{\o}we} and M. {Galili} and D. {Zibar}},
journal={Journal of Lightwave Technology},
title={Constellation Shaping for WDM Systems Using 256{QAM}/1024{QAM} With Probabilistic Optimization},
year={2016},
volume={34},
number={22},
pages={5146-5156},
doi={10.1109/JLT.2016.2607798},
ISSN={1558-2213},
}

@article{RennerJLT17,
author = {Julian Renner and Tobias Fehenberger and Metodi P. Yankov and Francesco {Da Ros} and S{\o}ren Forchhammer and Georg B{\"o}cherer and Norbert Hanik},
journal={Journal of Lightwave Technology},
number = {22},
pages = {4871--4879},
publisher = {OSA},
title = {Experimental Comparison of Probabilistic Shaping Methods for Unrepeated Fiber Transmission},
volume = {35},
year = {2017}
}

@article{BuchaliJLT16,
author = {Fred Buchali and Fabian Steiner and Georg B{\"o}cherer and Laurent Schmalen and Patrick Schulte and Wilfried Idler},
journal={Journal of Lightwave Technology},
number = {7},
pages = {1599--1609},
publisher = {OSA},
title = {Rate Adaptation and Reach Increase by Probabilistically Shaped 64-{QAM}: An Experimental Demonstration},
volume = {34},
year = {2016},
url = {http://jlt.osa.org/abstract.cfm?URI=jlt-34-7-1599}
}

@article{TuritsynOptica17,
author = {Sergei K. Turitsyn and Jaroslaw E. Prilepsky and Son Thai Le and Sander Wahls and Leonid L. Frumin and Morteza Kamalian and Stanislav A. Derevyanko},
journal = {Optica},
number = {3},
pages = {307--322},
publisher = {OSA},
title = {Nonlinear Fourier transform for optical data processing and transmission: advances and perspectives},
volume = {4},
year = {2017},
url = {http://www.osapublishing.org/optica/abstract.cfm?URI=optica-4-3-307},
doi = {10.1364/OPTICA.4.000307}
}

@Article{LeNatPhot2017,
author={Le, Son Thai
and Aref, Vahid
and Buelow, Henning},
title={Nonlinear signal multiplexing for communication beyond the Kerr nonlinearity limit},
journal={Nature Photonics},
year={2017},
volume={11},
number={9},
pages={570-576},
issn={1749-4893},
doi={10.1038/nphoton.2017.118},
url={https://doi.org/10.1038/nphoton.2017.118}
}

@ARTICLE{DongPTL15,
author={Z. {Dong} and S. {Hari} and T. {Gui} and K. {Zhong} and M. I. {Yousefi} and C. {Lu} and P. A. {Wai} and F. R. {Kschischang} and A. P. T. {Lau}},
journal={IEEE Photonics Technology Letters},
title={Nonlinear Frequency Division Multiplexed Transmissions Based on NFT},
year={2015},
volume={27},
number={15},
pages={1621-1623},
doi={10.1109/LPT.2015.2432793},
ISSN={1941-0174},
}

@ARTICLE{DaRosJLT19,
author={F. {Da Ros} and S. {Civelli} and S. {Gaiarin} and E. P. {da Silva} and N. {De Renzis} and M. {Secondini} and D. {Zibar}},
journal={Journal of Lightwave Technology},
title={Dual-Polarization NFDM Transmission With Continuous and Discrete Spectral Modulation},
year={2019},
volume={37},
number={10},
pages={2335-2343},
doi={10.1109/JLT.2019.2904102},
ISSN={1558-2213},
}

@INPROCEEDINGS{GaiarinCLEO20,
author={S. {Gaiarin} and R. {Jones} and F. {Da Ros} and D. {Zibar} },
booktitle = {Conference on Lasers and Electro-Optics (CLEO)},
journal = {Conference on Lasers and Electro-Optics},
title={End-to-end optimized nonlinear Fourier transform-basedcoherent communications},
publisher = {Optical Society of America},
year={2020},
pages = {SF2L.4},
}

@inproceedings{glorot2010understanding,
  title={Understanding the difficulty of training deep feedforward neural networks},
  author={Glorot, Xavier and Bengio, Yoshua},
  booktitle={Proceedings of the thirteenth international conference on artificial intelligence and statistics},
  pages={249--256},
  year={2010}
}

@inproceedings{dozat2016incorporating, % FIXME
  title={{Incorporating Nesterov momentum into Adam}},
  author={Dozat, Timothy},
  booktitle={International Conference on Learning Representations (ICLRW)},
  pages={1--6},
  year={2016}
}

@ARTICLE{BoscoPTL00, 
author={G. {Bosco} and A. {Carena} and V. {Curri} and R. {Gaudino} and P. {Poggiolini} and S. {Benedetto}}, 
journal={IEEE Photonics Technology Letters}, 
title={Suppression of spurious tones induced by the split-step method in fiber systems simulation}, 
year={2000}, 
volume={12}, 
number={5}, 
pages={489-491},
}

@article{baydin2017automatic,
  title={Automatic differentiation in machine learning: a survey},
  author={Baydin, At{\i}l{\i}m G{\"u}nes and Pearlmutter, Barak A and Radul, Alexey Andreyevich and Siskind, Jeffrey Mark},
  journal={The Journal of Machine Learning Research},
  volume={18},
  number={1},
  pages={5595--5637},
  year={2017},
  publisher={JMLR. org}
}

@inproceedings{li2018visualizing,
  title={Visualizing the loss landscape of neural nets},
  author={Li, Hao and Xu, Zheng and Taylor, Gavin and Studer, Christoph and Goldstein, Tom},
  booktitle={Advances in Neural Information Processing Systems},
  pages={6389--6399},
  year={2018}
}

@article{oliari2020regular,
  title={Regular perturbation on the group-velocity dispersion parameter for nonlinear fibre-optical communications},
  author={Oliari, Vin{\'\i}cius and Agrell, Erik and Alvarado, Alex},
  journal={Nature communications},
  volume={11},
  number={1},
  pages={1--11},
  year={2020},
  publisher={Nature Publishing Group}
}


%% file: bibliography-mendely.bib
@inproceedings{Klambauer2017,
author = {Klambauer, G{\"{u}}nter and Unterthiner, Thomas and Mayr, Andreas},
booktitle = {Advances in Neural Information Processing Systems},
isbn = {978-3-89937-157-4},
pages = {971--980},
title = {{Self-Normalizing Neural Networks}},
year = {2017}
}

@article{Yousefi2014p2,
author = {Yousefi, Mansoor I. and Kschischang, Frank R.},
doi = {10.1109/TIT.2014.2321151},
issn = {0018-9448},
journal = {IEEE Transactions on Information Theory},
number = {7},
pages = {4329--4345},
title = {{Information Transmission Using the Nonlinear Fourier Transform, Part II: Numerical Methods}},
url = {http://arxiv.org/abs/1204.0830 http://ieeexplore.ieee.org/document/6808508/},
volume = {60},
year = {2014}
}

@article{JonesPTL2018,
author = {Jones, Rasmus T. and Gaiarin, Simone and Yankov, Metodi P. and Zibar, Darko},
doi = {10.1109/LPT.2018.2831693},
issn = {1041-1135},
journal = {IEEE Photonics Technology Letters},
number = {12},
pages = {1079--1082},
title = {{Time-Domain Neural Network Receiver for Nonlinear Frequency Division Multiplexed Systems}},
url = {https://ieeexplore.ieee.org/document/8352849/},
volume = {30},
year = {2018}
}

@inproceedings{Li2018,
author = {Li, Shen and Hager, Christian and Garcia, Nil and Wymeersch, Henk},
booktitle = {European Conference on Optical Communication (ECOC 2018)},
doi = {10.1109/ECOC.2018.8535456},
isbn = {978-1-5386-4862-9},
number = {1},
pages = {1--3},
title = {{Achievable Information Rates for Nonlinear Fiber Communication via End-to-end Autoencoder Learning}},
url = {http://arxiv.org/abs/1804.07675 https://ieeexplore.ieee.org/document/8535456/},
year = {2018}
}

@article{Karanov2019a,
author = {Karanov, Boris and Lavery, Domani{\c{c}} and Bayvel, Polina and Schmalen, Laurent},
doi = {10.1364/OE.27.019650},
issn = {1094-4087},
journal = {Optics Express},
number = {14},
pages = {19650},
title = {{End-to-end optimized transmission over dispersive intensity-modulated channels using bidirectional recurrent neural networks}},
url = {https://www.osapublishing.org/abstract.cfm?URI=oe-27-14-19650},
volume = {27},
year = {2019}
}

@article{JonesECOC2018,
abstract = {A new geometric shaping method is proposed, leveraging unsupervised machine learning to optimize the constellation design. The learned constellation mitigates nonlinear effects with gains up to 0.13 bit/4D when trained with a simplified fiber channel model.},
author = {Jones, Rasmus T. and Eriksson, Tobias A. and Yankov, Metodi P. and Zibar, Darko},
doi = {10.1109/ECOC.2018.8535453},
file = {:media/work/bibliography/mendeley-pdf/2018/Jones et al. - Deep Learning of Geometric Constellation Shaping Including Fiber Nonlinearities - European Conference on Optical Communic.pdf:pdf},
isbn = {9781538648629},
journal = {European Conference on Optical Communication (ECOC)}, %IEEE Photonics Technology Letters
pages = {1--3},
publisher = {IEEE},
title = {{Deep Learning of Geometric Constellation Shaping Including Fiber Nonlinearities}},
year = {2018}
}

@article{Karanov2018,
author = {Karanov, Boris and Chagnon, Mathieu and Thouin, Felix and Eriksson, Tobias A. and Bulow, Henning and Lavery, Domanic and Bayvel, Polina and Schmalen, Laurent},
doi = {10.1109/JLT.2018.2865109},
issn = {0733-8724},
journal = {Journal of Lightwave Technology},
number = {20},
pages = {4843--4855},
title = {{End-to-End Deep Learning of Optical Fiber Communications}},
url = {https://ieeexplore.ieee.org/document/8433895/},
volume = {36},
year = {2018}
}

@article{Ablowitz2004a,
author = {Ablowitz, M J and Prinari, B and Trubatch, A D},
journal = {Dynamics of PDE},
number = {3},
pages = {239--299},
title = {{Integrable Nonlinear Schr{\"{o}}dinger Systems and their Soliton Dynamics}},
volume = {1},
year = {2004}
}

@article{Gaiarin2018,
author = {Gaiarin, S. and Perego, A. M. and da Silva, E. P. and {Da Ros}, F. and Zibar, D.},
doi = {10.1364/OPTICA.5.000263},
issn = {2334-2536},
journal = {Optica},
number = {3},
pages = {263},
title = {{Dual-polarization nonlinear Fourier transform-based optical communication system}},
url = {https://www.osapublishing.org/abstract.cfm?URI=optica-5-3-263},
volume = {5},
year = {2018}
}

@article{OShea2018,
author = {O'Shea, Timothy and Hoydis, Jakob},
doi = {10.1109/TCCN.2017.2758370},
isbn = {0423731238},
issn = {2332-7731},
journal = {IEEE Transactions on Cognitive Communications and Networking},
number = {4},
pages = {563--575},
pmid = {24432711},
title = {{An Introduction to Deep Learning for the Physical Layer}},
url = {http://ieeexplore.ieee.org/document/8054694/},
volume = {3},
year = {2017}
}

@article{Yousefi2014,
archivePrefix = {arXiv},
arxivId = {1202.3653},
author = {Yousefi, Mansoor I. and Kschischang, Frank R.},
doi = {10.1109/TIT.2014.2321143},
issn = {00189448},
journal = {IEEE Transactions on Information Theory},
number = {7},
pages = {4312--4328},
shorttitle = {Information Theory, IEEE Transactions on},
title = {{Information transmission using the nonlinear fourier transform, part I: Mathematical tools}},
url = {http://ieeexplore.ieee.org/lpdocs/epic03/wrapper.htm?arnumber=6808480},
volume = {60},
year = {2014}
}

@article{JonesArxiv2018,
archivePrefix = {arXiv},
eprint = {1810.00774},
author = {Jones, Rasmus T. and Eriksson, Tobias A. and Yankov, Metodi P. and Puttnam, Benjamin J. and Rademacher, Georg and Luis, Ruben S. and Zibar, Darko},
pages = {1--9},
title = {{Geometric Constellation Shaping for Fiber Optic Communication Systems via End-to-end Learning}},
url = {http://arxiv.org/abs/1810.00774},
year = {2018}
}

@article{Gaiarin2018PTL,
author = {Gaiarin, Simone and {Da Ros}, Francesco and {De Renzis}, Nicola and {Da Silva}, Edson P. and Zibar, Darko},
doi = {10.1109/LPT.2018.2874204},
issn = {10411135},
journal = {IEEE Photonics Technology Letters},
number = {22},
pages = {1983--1986},
title = {{Dual-Polarization NFDM Transmission Using Distributed Raman Amplification and NFT-Domain Equalization}},
volume = {30},
year = {2018}
}

@article{Chen2016,
archivePrefix = {arXiv},
arxivId = {1604.06174},
author = {Chen, Tianqi and Xu, Bing and Zhang, Chiyuan and Guestrin, Carlos},
pages = {1--12},
title = {{Training Deep Nets with Sublinear Memory Cost}},
url = {http://arxiv.org/abs/1604.06174},
year = {2016}
}

@misc{MemorySavingGradient,
  author = {},
  title = {Gradient checkpoint},
  publisher = {GitHub},
  journal = {GitHub repository},
  howpublished = {\url{https://github.com/cybertronai/gradient-checkpointing}},
  commit = {4f57d6a0e4c030202a07a60bc1bb1ed1544bf679}
}


%% file: bibliography-revision.bib
@inproceedings{wahls2017generation,
  title={Generation of time-limited signals in the nonlinear Fourier domain via b-modulation},
  author={Wahls, Sander},
  booktitle={2017 European Conference on Optical Communication (ECOC)},
  pages={1--3},
  year={2017},
  organization={IEEE}
}

@article{musumeci2018overview,
  title={An overview on application of machine learning techniques in optical networks},
  author={Musumeci, Francesco and Rottondi, Cristina and Nag, Avishek and Macaluso, Irene and Zibar, Darko and Ruffini, Marco and Tornatore, Massimo},
  journal={IEEE Communications Surveys \& Tutorials},
  volume={21},
  number={2},
  pages={1383--1408},
  year={2018},
  publisher={IEEE}
}

@article{khan2019optical,
  title={An optical communication's perspective on machine learning and its applications},
  author={Khan, Faisal Nadeem and Fan, Qirui and Lu, Chao and Lau, Alan Pak Tao},
  journal={Journal of Lightwave Technology},
  volume={37},
  number={2},
  pages={493--516},
  year={2019},
  publisher={IEEE}
}

@article{GaiarinJLT2020,
  title={Experimental demonstration of nonlinear frequency division multiplexing transmission with neural network receiver},
  author={Gaiarin, Simone and DaRos, Francesco and De Renzis, Nicola and Jones, Rasmus Thomas and Zibar, Darko},
  journal={Journal of Lightwave Technology},
  year={2020},
  publisher={IEEE}
}
